\documentclass[review,3p,11pt, sort&compress]{elsarticle}

\usepackage{our_commands}

\allowdisplaybreaks
\journal{ }

\begin{document}

\begin{frontmatter}

\title{A kinetic model of a polyelectrolyte gel undergoing phase separation}

\author[1]{Giulia~L.~Celora\corref{cor1}}
\ead{celora@maths.ox.ac.uk}

\author[1]{Matthew G.~Hennessy}
\ead{hennessy@maths.ox.ac.uk}

\author[1]{Andreas~M\"unch}
\ead{muench@maths.ox.ac.uk}

\author[2]{Barbara~Wagner}
\ead{Barbara.Wagner@wias-berlin.de}

\author[1]{Sarah~L.~Waters}
\ead{waters@maths.ox.ac.uk}

\cortext[cor1]{Corresponding author}

\address[1]{Mathematical Institute, Woodstock Road, University of Oxford, Oxford, OX2 6GG, UK}
\address[2]{Weierstrass Institute, Mohrenstrasse 39, 10117 Berlin, Germany}

\begin{abstract}
In this study we use non-equilibrium thermodynamics to systematically derive a phase-field model of a polyelectrolyte gel coupled to a hydrodynamic model for a salt solution surrounding the gel. The 
governing equations for the gel account for the free energy of the internal interfaces which form upon phase separation, the nonlinear elasticity
of the polyelectrolyte network, and
multi-component diffusive transport following a Stefan--Maxwell approach. 
The time-dependent model describes the evolution of the gel across multiple time and spatial scales and so is able to capture the large-scale solvent flux and the emergence of long-time pattern formation in the system.
We explore the model for the case of a constrained gel undergoing uni-axial deformations. Numerical simulations show that rapid changes in the gel volume
occur once
the volume phase transition sets in, as well as the triggering of spinodal decomposition
that leads to strong inhomogeneities in the lateral stresses,
potentially leading to experimentally visible patterns.
\end{abstract}


\end{frontmatter}

\section{Introduction}
A polyelectrolyte gel is a network of covalently cross-linked polyelectrolyte macromolecules that are swollen with a fluid. The polyelectrolyte chains are electrically charged and they interact with dissolved ions in the imbibing fluid. 
If placed in a salt solution,  hereafter referred to as an {\it ionic bath}, polyelectrolyte gels undergo chemical, electrical, and mechanical
interactions which trigger 
structure formation in the gel.
The exchange of solvent and mobile ions between the ionic bath and the polyelectrolyte gel enables the system to be driven from one equilibrium state to another. The transition between equilibria involves a volume phase transition in the form of swelling or collapse of the gel \cite{matsuo_patterns_1992,zubarev_self-similar_2004} and has been widely discussed both experimentally and theoretically \cite{Tanaka1978, Dusek1968, dobrynin_theory_2008,Dimitriyev_2019,mccoy_dynamic_2010,mussel_experimental_2019,Horkay2001, Bertrand2016}. 
This process depends in a subtle way on the concentration and valency of the salt in the solvent, the material properties of the macromolecules including the degree of fixed charge, and externally applied fields, such as temperature or an electric field. When considering the transient evolution of the gel between the different equilibria, phase separation can occur whereby regions of highly swollen and collapsed gel co-exist. As shown experimentally by Tanaka et al.~\cite{matsuo_patterns_1992,Tanaka1987}, phase separation gives rise to surface instabilities which can transiently or permanently affect the gel morphology and its resulting properties. Moreover, small changes in the salt concentration in the surrounding ionic bath has dramatic effects on the gel state and can result in phenomena such as discontinuous phase transitions connected with super-collapse \cite{Khokhlov_PolyelecCollape_1996,hua_theory_2012} or re-entrant swelling \cite{sing_reentrant_swelling_2013}. 
Environmental changes of the gel may 
also result in  micro- or nanophase separation \cite{shibayama_spatial_1998, wu_control_2010}
or reverse Ostwald ripening \cite{rosowski_elastic_2020, rosowski_elastic_2020-1}, features that also play a role in biological processes such as in subcellular organelle formation \cite{Brangwynne2015, style_liquidliquid_2018}. 
In regenerative medicine, polyelectrolyte gels can be used as scaffolds for cultures of biological cells to engineer replacement biological tissues \cite{Kwon2014, Lutolf2005, Ning2018}. The local electromechanical environment experienced by cells within these polyelectrolyte scaffolds can be controlled by the application of an external electric field, so that these bio-active materials can be utilised to guide and direct cell behaviour.  Polyelectrolyte gels are also used as model systems for biological tissues such as cartilage and the brain \cite{Lutolf2005, Lang2014}. 
Phase separation of environmentally sensitive polyelectrolyte gels is also fundamental to the development of smart, responsive materials and sensors \cite{Buenger2012,Chaterji2007,Hong2012,Stuart2010}.  

To capture the underlying physics, which combines aspects of electrochemistry and condensed matter physics, we use non-equilibrium thermodynamics to systematically derive a phase-field model of a polyelectrolyte gel that accounts for the free energy of the internal interfaces which form upon phase separation, as well as for finite elasticity and electrostatic interactions between charged species. We also derive a thermodynamically consistent model for the ionic bath. In contrast to the polyelectrolyte gel, the ionic bath does not have a well-defined reference configuration, which leads to subtle differences when using our thermodynamic framework to obtain constitutive relationships.  The governing equations for the bath that we derive
are similar to those obtained for liquid electrolytes~\cite{Roubivcek2006, Dreyer2015} used in electrochemical storage systems such as batteries. Our model accounts for multi-component transport of solvent and mobile ions using a Stefan--Maxwell approach. By accounting for cross-diffusion, we avoid anomalous diffusivities which arise when the underlying theory assumes that the diffusive flux of a species is solely driven by the gradient of its own chemical potential \cite{Krishna1997}. The Stefan--Maxwell approach \cite{Stefan1871,Maxwell1867} correctly captures the hydrodynamic drag between different components of the mixture by balancing the friction forces between the different species \cite{bothe2020structure}. While having been previously ignored for polyelectrolyte gels \cite{Hong2010}, the recent work by Zhang et al.~\cite{Zhang2020} has highlighted the role of cross-diffusion in modelling effects such as a temporary excess of salt entering the hydrogel during swelling, which is subsequently rejected as the gel approaches its new equilibrium.  

Near the interface between a polyelectrolyte gel and a salt bath, ions from
the bath accumulate in order to screen the electric charges on the
polyelectrolyte macromolecules. This gives rise to a diffuse layer of charge
known as an electric double layer. The amount of charge in the electric double
layer typically decreases with the distance from the gel-bath interface until
the gel and the bath become electrically neutral. 
Mori et al.~\cite{Mori2013} derived a coupled model for a polyelectrolyte gel in contact with an ionic bath and considered the electroneutral limit in which the thickness of the electric double layer tends to zero.
However, they did not consider phase separation in the gel nor Stefan--Maxwell diffusion. The model presented here can
resolve the structure of the electric double layer and provide a new
understanding of the role it plays during dynamic phase transitions. 
We explore this in detail in a companion paper \cite{Hennessy_debye_2020},
where we also use matched asymptotic expansions to derive the corresponding electroneutral model for the
gel-bath system with consistent jump conditions to be imposed across the
electric double layer.

The model presented in this paper thus captures the fully time-dependent hydrodynamics of the ionic bath, and the electro-chemo-mechanical interactions between the gel and the ionic bath. The model can be exploited to determine the fast dynamics of ion migration and its impact on the much slower transport of the solvent, which in turn provides mechanistic insights into the  transient dynamics of the emerging patterns within the gel en route to a new equilibrium state. This theoretical framework enables new interpretations of experimental results as it can be used as a tool to discover and investigate the phenomena of phase separation in polyelectrolyte gels. 

In \S 2, we derive the model  for the polyelectrolyte gel by formulating the conservation laws, constructing the free energies, and using the approach by Gurtin \cite{Gurtin1996,Gurtin2010book} to derive the constitutive equations via an energy imbalance inequality. The full model for the ionic bath is derived in \S 3 using a similar approach. The interfacial conditions between the gel and ionic bath are given in \S4. 
Some first insights into the potential of the model to predict the emergence of patterns in the gel are discussed in \S 5,  where we explore numerically the dynamics of swelling and collapse of a constrained gel. We use our model to resolve the transient dynamics of the volume phase transition. In some cases, this transition is also accompanied by the emergence of locally collapsed states that arise due to spinodal decomposition ahead of the main transition front. Interestingly, the anion concentration is higher in the collapsed state compared to the swollen state, whereas the cation fraction is lower. Moreover, spinodal decomposition gives rise to a highly non-uniform state of stress, suggesting the emergence of patterns that may be observed in experiments. The numerical results presented here are extended in a companion paper \cite{Celora_siap_2020}, where we also carry out a linear stability analysis to identify the parameter regimes where spinodal decomposition occurs together with a phase-plane analysis to elucidate the mathematical structure of the propagating transition fronts. 

\section{Model derivation for a polyelectrolyte gel}
In this section we derive the governing equations describing the behaviour of a polyelectrolyte gel. In \S\ref{ModelSetUp} we describe the components of the polyelectrolyte gel, and introduce the reference (dry) and current (swollen) configurations. In \S \ref{conslaws} we present the equations representing incompressibility of the polyelectrolyte gel, conservation of the mobile species, and conservation of momentum of the gel. These are complemented with the definition of the electrostatic potential and Gauss' law of electrostatics for the electric displacement. The equations are presented in both the reference and the current configurations. In \S\ref{free_energy} we construct the free energy of the system. We then apply the energy imbalance inequality of Gurtin \cite{Gurtin2010book} to obtain thermodynamically consistent expressions for the constitutive equations in terms of the previously specified free energy (\S\ref{sec_ine}).~In \S\ref{ent} we consider the rate of entropy production to determine relationships for the mass fluxes. Finally in \S\ref{gelcurrent} we present the full model equations (governing equations together with constitutive equations) in the current state. 

\subsection{Kinematics} \label{ModelSetUp}

\begin{figure}[htb]
	\centering
	\includegraphics[width=0.6\textwidth]{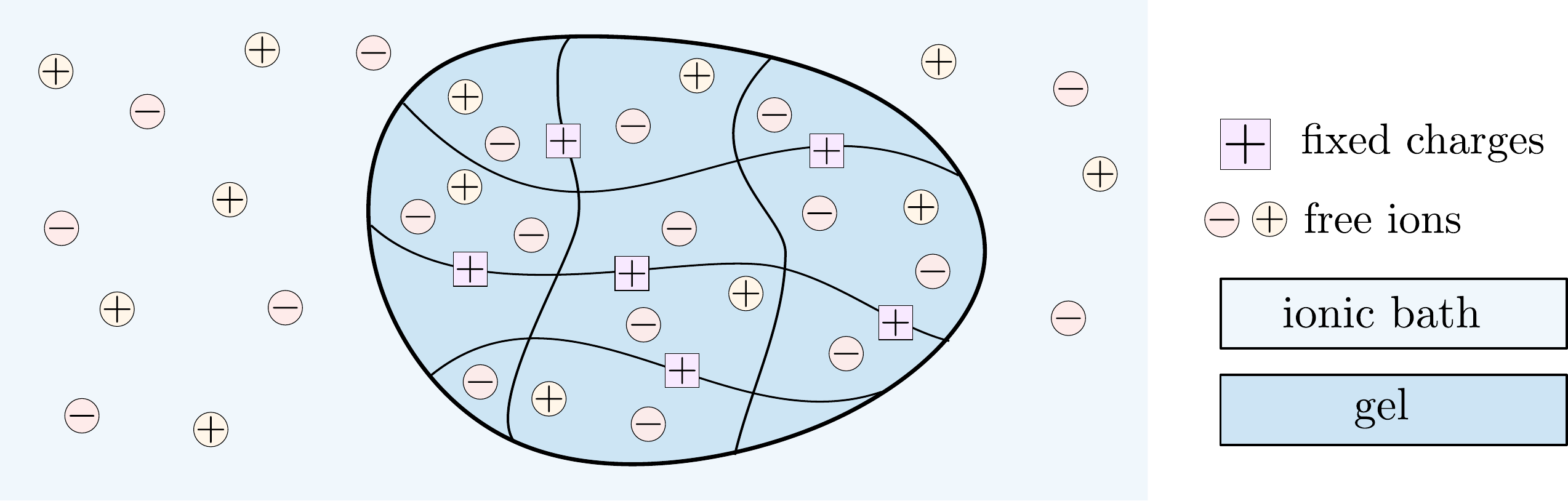}
	\caption{Schematic representation of a polyelectrolyte gel in contact with an ionic bath.  The gel is a three-phase material, composed of solid polymer network with fixed charges, solvent, and freely moving ions. The ionic bath consists of solvent and freely moving ions.}
	\label{gel_schem}
\end{figure}
We consider the gel as a multi-phase material composed of a solid polymer network with fixed charges and a solution consisting of solvent, such as water, and $N$ freely moving ionic species (\textit{i.e.}~solutes), see Figure \ref{gel_schem}. We assume all phases are intrinsically incompressible and isotropic.  We note that the solid-phase volume  encompasses the fixed-charge volume. Throughout the model derivation, we use subscripts $n, s$ to denote the solid polymer network and solvent respectively, the index $i=1,\ldots,N$ to denote the ionic species, and the index $m\in\left\{s,1,\ldots,N\right\}$ to refer to the species that are mobile relative to the network, {\it i.e.}~both the solvent and solutes. For later convenience, we introduce the set notation $\mathbb{I}=\{1, \ldots, N \}$ and $\mathbb{M}=\{s,1, \ldots, N \}$.
\begin{figure}[h!]
	\centering
	\includegraphics[width=0.85\textwidth]{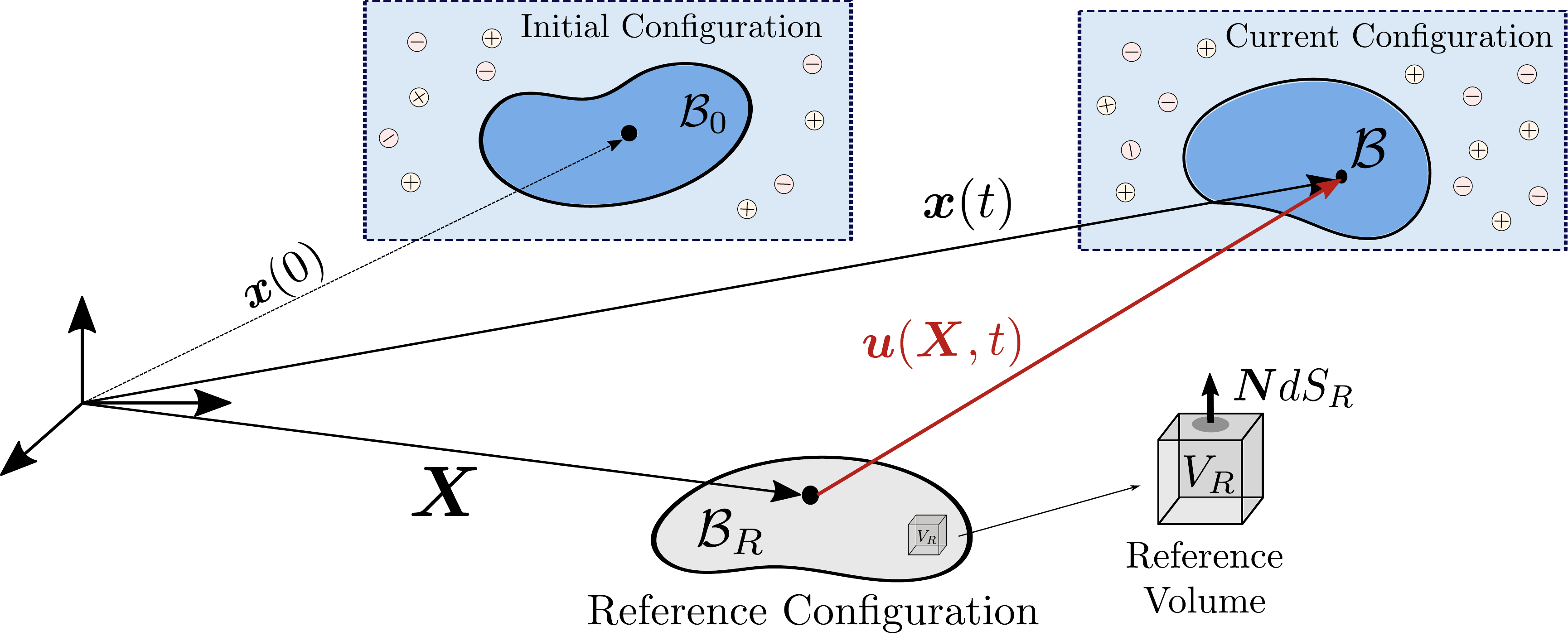}
	\caption{Sketch of the reference, initial and current state of the gel.}
	\label{gel_configuration}
\end{figure}

The motivation for this study is to understand the swelling behaviour of a polyelectrolyte gel in contact with an ionic bath. By changing the conditions in the bath, such as the concentration of ions, the polyelectrolyte gel will swell or shrink. To capture this behaviour, we consider a gel that is initially pre-swollen and in equilibrium with the surrounding bath. We then alter the conditions
in the bath to drive the gel towards a new equilibrium state. 
As shown in Figure \ref{gel_configuration}, the initial configuration differs from the reference state, the latter of which is stress free and assumed to be the dry gel, \textit{i.e.} only the solid phase is present. As the gel deforms, the material element located at $\vec{X}$ (Lagrangian coordinates) in the reference configuration $\mathcal{B}_R$ is displaced to the point $\vec{x}$ (Eulerian coordinates) in the current configuration $\mathcal{B}$ as shown in  Figure \ref{gel_configuration}. Such a transformation is described by the deformation gradient tensor $\tens{F}= \partial \vec{x}/\partial \vec{X}$; information about the change in volume during deformation is encoded in $J= \det \tens{F}$, while $\vec{u}= \vec{x}-\vec{X}$ is the displacement vector. We move from the reference state to current configuration as follows. Given a volume element $dV_R$ and its surface element $\vec{N}dS_R$ in the reference configuration, these are related to the corresponding  quantities $dV$ and $\vec{n}dS$ in the current configuration by the relations:
\begin{equation}
dV=JdV_R,\quad \vec{n}dS=J \tens{F}^{-T}\vec{N}dS_R,\label{elem}
\end{equation}
where $\vec{N}$ and $\vec{n}$ are respectively the normal unit vector to the surface elements $dS_R$ and $dS$ respectively, see Figure~\ref{gel_configuration}. 

As the gel deformation is determined by the displacement of the solid phase, the solid phase velocity in the current configuration, $\vec{v}_n$, and displacement are related, so that $\vec{v}_n=\dot{\vec{u}}$ where dots denote derivatives with respect to time in the reference configuration, {\it i.e.}~$\left.{\partial}/{\partial t}\right |_{\vec{X}} \vec{u}$. Equivalently, in the current configuration, we can relate the network velocity to the displacement via $\vec{v}_n=\partial\vec{u}/\partial t +(\vec{v}_n\cdot\nabla)\vec{u}$, where $\nabla$ is the gradient with respect to $\vec{x}$.

\subsection{Conservation equations and electrostatics} \label{conslaws}

\paragraph*{Reference configuration} As the solid phase is incompressible, any change in the volume during deformation is due to the migration of solvent and solute molecules, whose nominal concentrations (\emph{i.e.} number of molecules per unit volume in $\mathcal{B}_R$) are denoted by $C_m$ ($m\in\mathbb{M}$). This leads to the molecular incompressibility condition:
\begin{subequations}
	\begin{align}
		J= 1 + \sum_{m\in\mathbb{M}} \nu_m C_m,
		\label{incomp}
	\end{align}
	where $\nu_m$ ($m\in\mathbb{M}$) is the molecular volume of each species in the solution.    
	
	Conservation of each mobile species in the reference configuration reads:
	\begin{align}
		\dot{C}_m + \nabla_R \cdot \vec{J}_m = 0, \label{consmass}
	\end{align}
	where $\vec{J}_m$ is the nominal flux per unit area in the reference state and $\nabla_R$ denotes the gradient with respect to the Lagrangian coordinates $\vec{X}$. 
	
	When considering gels, inertial and gravitational effects are commonly neglected, so that the conservation of momentum for the gel reads:
	\begin{align}
		\nabla_R \cdot \tens{S}=\mathbf{0} \label{consmom},
	\end{align}
	where $\tens{S}$ is the first Piola--Kirchhoff tensor, which represents the stress state of the polyelectrolyte gel in the reference configuration.
	
	The accumulation of electric charge generates an electric field which is denoted by $\vec{E}$ in the reference configuration. By introducing the electrostatic potential $\Phi$, we have that:
	\begin{align}
		\vec{E}= -\nabla_R \, \Phi, 
		\label{Phi}
	\end{align}
	As is common in the modelling of polyelectrolyte gels \cite{Drozdov2015,Drozdov2015b,Hong2012,Yu2017}, we consider the gel to be a linear dielectric material. Consequently, the presence of the electric field generates an electric displacement, $\vec{H}$, which must obey Gauss' law of electrostatics:
	\begin{align}
		\nabla_R \cdot \vec{H}= Q=e\left(\sum\limits_{i\in\mathbb{I}} z_i C_i+z_f C_{f}\right),
		\label{gauss}
	\end{align}
	where $Q$ is the nominal total charge density, which accounts for both fixed and moving charges; $e$ is the elementary charge; $C_f$ is the nominal concentration of fixed charges; and $z_i$ is the valence of the corresponding charged species. 
\end{subequations}

\paragraph*{Current Configuration}
By integrating the governing equations in the reference state over an arbitrary control volume $\mathcal{V}_R$ and applying~(\ref{elem}), we can easily derive the corresponding integral form of the governing equations for an arbitrary control volume $\mathcal{V}$. From these it is straightforward to determine the local form of the equation appealing to the fact that $\mathcal{V}$ is arbitrary.
Since these arguments are standard, we present only the final result, \emph{i.e.} the local form of the governing equations in the current configuration.
However, in \ref{conscurrent}, the full calculation is shown for the equations
governing the concentrations.

The concentration of mobile species in the current configuration, i.e. molecules per unit volume in $\B$, is given by $c_m=C_m/J$. The incompressibility condition (\ref{incomp}) then reads
\begin{subequations}
	\begin{align}
		J=\frac{1}{1-\sum_{m\in\M} \nu_m c_m}.\label{incompcurrent}
	\end{align}
	Conservation of mobile species is given by
	\begin{align}
		\partial_t{c}_m + \nabla \cdot (c_m \vec{v}_n)=-\nabla \cdot \vec{j}_m, \label{consmass_current}
	\end{align}
	where $\vec{j}_m=J^{-1}\tens{F}\vec{J}_m$ is the diffusive flux of the $m$-th species.
	We define the velocity of the mobile species to be $\vec{v}_m$, and relate it to the flux $\vec{j}_m$ as follows:
	\begin{align}
		\vec{j}_m=c_m \left(\vec{v}_m-\vec{v}_n\right)=c_m\bar{\vec{v}}_{m},\label{flux_def}
	\end{align}
	where, for later convenience, we introduce the notation $\bar{\vec{v}}_{m}$ for the relative velocity of the $m$-th phase with respect to the network velocity. 
	
	When considering conservation of momentum (\ref{consmom}), the counterpart to the first Piola--Kirchhoff tensor $\tens{S}$ in the current configuration is the Cauchy stress tensor $\tens{T}$, which is 
	$\tens{T} = J^{-1}\tens{S}\tens{F}^T$.
	Conservation of momentum is given by
	\begin{align}
		\nabla \cdot \tens{T}=\mathbf{0} \label{consmomcurrent}.
	\end{align}
	Finally, following~\cite{Drozdov2015b}, we have that $\vec{h}=J^{-1}\tens{F} \vec{H}$ and $\vec{e}=\tens{F}^{-T}\vec{E}$ are respectively the (current) electric displacement and field. We can therefore reformulate equations~(\ref{Phi})-(\ref{gauss}) as: 
	\begin{align}
		\vec{e}&=-\nabla \Phi,\label{phicurrent}\\
		\nabla \cdot \vec{h}&=q= \frac{Q}{J}=e\left(\sum\limits_{i\in\mathbb{I}} z_i c_i+z_f c_{f}\right).\label{gausscurrent}
	\end{align}
\end{subequations}
\subsection{Free energy}\label{free_energy}
Equations of state that are consistent with the second law of thermodynamics can be derived by specifying the precise form of the free energy per unit volume in the reference configuration $\Psi$.  We assume that the free energy is composed of five contributions as follows
\begin{align}
	\Psi=\Psi_1+\Psi_2+\Psi_3+\Psi_4+\Psi_5,\label{free}
\end{align}
corresponding to the energy of the electric field ($\Psi_1$); the energy of the mobile species not interacting with each other or the network ($\Psi_2$); the energy of mixing different chemical species ($\Psi_3$); the interfacial energy between dissimilar phases ($\Psi_4$); and the the elastic energy of the gel ($\Psi_5$).

Assuming the polyelectrolyte gel to be an ideal and linear dielectric material, with constant permittivity $\epsilon$, the free energy of polarisation is given by \cite{Drozdov2015,Drozdov2015b,Hong2012}:
\begin{equation}
\Psi_1 = \frac{1}{2\epsilon J} (\tens{F}\vec{H}) \cdot \tens{F} \vec{H}.
\end{equation}
We assume that the permittivity is dominated by the permittivity of the solvent, with little contribution from the network or mobile phases. 

The second energy density $\Psi_2$ has the standard form \cite{Drozdov2016b}:
\begin{equation}
\Psi_2 = \sum\limits_{m\in\mathbb{M}} \mu^0_m C_m
\end{equation} 
where $\mu^0_m$ denotes the chemical potential of non-interacting mobile species.

When considering the mixing energy, a common assumption in the study of polyelectrolyte gels is that the leading contribution to the enthalpy arises from the solvent due to the hydrophobic interaction with the solid phase \cite{Hua2012}. According to the Flory--Huggins theory \cite{Flory1942,Huggins1942} of mixtures, the mixing energy $\Psi_3$ is then given by:
\begin{equation}
\Psi_3 = k_B T \left(\frac{ \chi C_s}{1+\sum_{j\in \M} \nu_j C_j}+\sum_{m\in\mathbb{M}} C_m \ln \frac{\nu_mC_m}{1+\sum_{j\in \M} \nu_j C_j}\right),\label{mix}
\end{equation}
where $k_B$ is Boltzmann's constant, $T$ is the temperature and $\chi$ is the Flory interaction parameter
(capturing the enthalpy of mixing the solvent and solid phase).

The effect of interfacial tension may be included by considering interfacial energy contributions. For multiple phases, and assuming ideal interfaces \cite{Hong2013}, candidate interfacial energy contributions expressed in the current configurations include \cite{Garcke_2004}
\begin{equation}
\psi_4=\sum\limits_{m\in \mathbb{S}} A_m \left|\nabla c_m\right|^2 \ \ \mbox{or} \ \ 
\psi_4=\sum\limits_{l,m\in \mathbb{S}} B_{lm}\left|c_l\nabla c_m-c_m\nabla c_l\right|^2,
\end{equation}
where $\mathbb{S}=\left\{ n\right\} \cup \mathbb{M}$
and the coefficients $A_m$ and $B_{lm} = B_{ml}$ are constants. Note that we here use $\psi$ to denote the free energy per unit volume of the \textit {current configuration}. 
While our model formulation can accommodate contributions to the interfacial energy arising from gradients in all concentrations, here we consider the case for dilute ion concentrations, and assume that the interfacial energy contribution is dominated by gradients in the solvent concentration only. In the current configuration, we consider the specific form of $\psi_4$ to be 
\begin{equation}
\psi_4 = \frac{\gamma}{2}\left|\nabla c_s\right|^2,\label{inteng}
\end{equation}
where $\gamma$ plays a role analogous to surface tension (see also \cite{Onuki1998}).  This interfacial energy expressed in a Cartesian representation of the
reference configuration is
\begin{equation}
\begin{aligned}
\Psi_4=\frac{\gamma}{2J}G_{iJ}G_{iM}\frac{\partial C_s}{\partial X_J}\frac{\partial C_s}{\partial X_M}+\frac{\gamma C_s^2}{2J^3}&G_{iJ}G_{iM}\frac{\partial J}{\partial X_J}\frac{\partial J}{\partial X_M}-\frac{\gamma C_s}{J^2}G_{iJ}G_{iM}\frac{\partial C_s}{\partial X_J}\frac{\partial J}{\partial X_M},
\end{aligned}
\end{equation}
where $\tens{G}=\tens{F}^{-T}$ and summation over repeated indices is implied.

Finally, for the strain energy we consider the gel to be a hyperelastic neo-Hookean material:
\begin{equation}
\Psi_5= \frac{G}{2} \left(\tens{F}:\tens{F} - 3 -2 \ln J\right)\
\end{equation}
where $G$ is the shear modulus.
\subsection{Energy imbalance inequality}\label{sec_ine}

As derived by Gurtin \cite{Gurtin1996,Gurtin2010book} when considering isothermal processes, the second law of thermodynamics can be rewritten in terms of the Helmholtz free energy $\Psi$, leading to the \textit{energy imbalance inequality}:
\begin{equation}
\frac{\d}{\d t} \left\{\int_{\mathcal{V}_R} \Psi dV_R\right\}\leq \mathcal{W}(\mathcal{V}_R) + \mathcal{M}(\mathcal{V}_R) \label{energyin}
\end{equation}
where $\mathcal{V}_R$ is an arbitrary control volume in the reference configuration $\mathcal{B}_R$,  $\mathcal{W}(\mathcal{V}_R)$ is the rate at which the surrounding environment does work on $\mathcal{V}_R$ and $\mathcal{M}(\mathcal{V}_R)$ is the rate of change of energy associated with the addition of mass due to transport processes.

The term $\mathcal{W}(\mathcal{V}_R)$ is decomposed into two contributions, the rate of electrical work, $W_{el}(\mathcal{V}_R)$, and the rate of mechanical work,  $\mathcal{W}_{mec}(\mathcal{V}_R)$. Following \cite{Drozdov2016}, $\mathcal{W}_{el}(\mathcal{V}_R)$ is defined as:
\begin{equation}
\mathcal{W}_{el}(\mathcal{V}_R) = -\int_{\mathcal{S}_R} \Phi\, \dot{\vec{H}}\cdot \vec{N}dS_R,\label{electricalwork}
\end{equation}
where $\mathcal{S}_R$ is the surface associated with the control volume $\mathcal{V}_R$.

To determine the rate of mechanical work, $\mathcal{W}_{mec}(\mathcal{V}_R)$, we follow Gurtin \cite{Gurtin1996} and account for the presence of both macro-stresses $\tens{S}$, and micro-stresses $\vec{\xi}$, the latter of which arise due to the heterogeneity of the system. More specifically, we assume that the micro-stresses originate from gradients in the concentration of mobile species, {\it i.e.}~solvent and ions. As the energy imbalance is formulated in the reference configuration $\mathcal{B}_R$, we replace $c_m$ by $C_m/J$ when computing composition gradients. The micro-stresses associated with gradients in $C_m$ and $J$ are denoted by $\vec{\xi}_m$ and $\vec{\xi}_J$, respectively. The total rate of mechanical work $W_{mec}(\mathcal{V}_R)$ then reads \cite{Gurtin1996}:
\begin{equation}	
\centering
\mathcal{W}_{mec}(\mathcal{V}_R) =\int_{\mathcal{S}_R} \tens{S}\vec{N} \cdot \dot{\vec{u}}\, dS_R+  \int_{\mathcal{S}_R} \left[\sum_{m\in \M} \left(\vec{\xi}_m\cdot \vec{N}\right)\dot{C}_m + \left(\vec{\xi}_J\cdot \vec{N}\right)\dot{J}\right]\, dS_R.\label{mechanicalwork}
\end{equation}

The system exchanges mass due to the diffusion of each mobile species, so $\mathcal{M}(\mathcal{V}_R)$ is given by:
\begin{equation}
\mathcal{M}(\mathcal{V}_R)= \sum\limits_{m\in \M} - \int_{\mathcal{S}_R} \mu_m \,\vec{J}_m \cdot \vec{N}\, dS_R\, , \label{mass-rate}
\end{equation}
where $\mu_m$ is the chemical potential associated with each species. 

Since equation (\ref{energyin}) must hold for arbitrary\footnote{We assume the control volume is not at the boundary between the polyelectrolyte gel and surrounding ionic solution.} control volumes $\mathcal{V}_R$, substituting equations (\ref{electricalwork}), (\ref{mechanicalwork}) and (\ref{mass-rate}) into the energy imbalance inequality~(\ref{energyin}), and applying the divergence theorem, we obtain the following localised inequality:
\begin{equation}
\begin{aligned}
\dot{\Psi}  + \, \sum\limits_{i\in\mathbb{I}} \left[e \Phi  z_i - \mu_i-\nabla_R \cdot \vec{\xi}_i\right] \dot{C}_i - (\mu_s + \nabla_R \cdot \vec{\xi}_s)\,\dot{C}_s-\sum_{m\in \M}\vec{\xi}_m \cdot \nabla_R \, \dot{C}_m\\
-\left(\tens{S} +J\left(\nabla_R \cdot \vec{\xi}_J\right)\tens{F}^{-T}\right):\dot{\tens{F}}
- \vec{E}\cdot \dot{\vec{H}} \, -\vec{\xi}_J \cdot \nabla_R \, \dot{J}
+ \sum\limits_{m\in\mathbb{M}} \nabla_R \, \mu_m \cdot \vec{J}_m \leq 0,
\label{temp2}
\end{aligned} 
\end{equation}
where additionally we have exploited equations~(\ref{consmass}) and (\ref{consmom}). We account for the constraint imposed by the incompressibility condition~(\ref{incomp}) by differentiating with respect to time to give
\begin{gather}
	\sum\limits_{m\in\mathbb{M}} \nu_m\dot{C}_m - J \tens{F}^{-T}:\dot{\tens{F}} =0, \label{temp3}
\end{gather}
and including the constraint~(\ref{temp3}) in (\ref{temp2}) via a Lagrange multiplier $p$, where the multiplier $p$ plays the role of the thermodynamic pressure. 

Following  \cite{Lallit2011, Gurtin1996} and guided by equation (\ref{temp2}), we consider the free energy to have the following dependencies
\begin{equation}
\Psi = \Psi (\tens{F}, C_m, \nabla_R \,C_m, \nabla_R \,J, \vec{H}). \label{psi_def}
\end{equation}
We highlight that we have retained the dependence of $\Psi$  on $\nabla_RC_m\, , m\in\mathbb{M}$, to make explicit the  relationship between  the microstresses $\vec{\xi}_m$ and $\nabla_RC_m$ in the discussion that now follows. 
However, in the specification of the forms of the constitutive relationships in \S\ref{free_energy}, the subsequent analysis will considerably simplify and we will retain only the microstresses $\vec{\xi}_s$ and $\vec{\xi}_J$.

Substituting~(\ref{psi_def}) into~(\ref{temp2}),  and including the constraint~(\ref{temp3}), we obtain the augmented form of the energy imbalance inequality:
\begin{equation}
\begin{aligned}
\sum_{m\in\M}\left(\frac{\partial \Psi}{\partial \nabla_R C_m}-\vec{\xi}_m\right) \cdot \nabla_R \dot{C}_m + \left(\frac{\partial \Psi}{\partial C_s}-\mu_s-\nabla_R \cdot \vec{\xi}_s+ \nu_s p \right)\dot{C}_s
+\left(\frac{\partial \Psi}{\partial \vec{H}}-\vec{E}\right) \cdot \dot{\vec{H}}\\ +\left(\frac{\partial \Psi}{\partial \nabla_R J}-\vec{\xi}_J\right)\cdot \nabla_R \dot{J}
+ \sum_{i\in\mathbb{I}}\left(\frac{\partial \Psi}{\partial C_i} + e\Phi z_i-\mu_i-\nabla_R \cdot \vec{\xi}_i+\nu_i p\right) \dot{C}_i\\
+  \left(\frac{\partial \Psi}{\partial \tens{F}} - \tens{S} - p J \tens{F}^{-T}-J(\nabla_R \cdot \vec{\xi}_J)\tens{F}^{-T}\right): \dot{\tens{F}}+ \sum_{m\in\mathbb{M}} \nabla_R \,\mu_m \cdot \vec{J}_m \leq 0, \label{ineq}
\end{aligned}
\end{equation}
Inequality (\ref{ineq}) is linear in $\nabla_R \dot{C}_m$, $\dot{C}_m$,  $\nabla_R \dot{J}$, $\dot{\vec{H}}$, $\dot{\tens{F}}$ and $\nabla_R\, \mu_m$, each of which can be chosen independently at each point $\vec{X}$ and time $t$  \cite{Gurtin1996}. In particular, we can choose $\nabla_R\, \mu_m=0$ and vary the other variables independently. The only way for the inequality~(\ref{ineq}) to be valid in all cases is to assume that the brackets are identically zero.  This gives the following constitutive equations
\begin{subequations}
	\begin{align}
		\vec{\xi}_J&=\frac{\partial \Psi}{\partial \nabla_R J},\qquad\vec{\xi}_m	=\frac{\partial \Psi}{\partial \nabla_R C_m}, \  m\in \M,\label{state_eq1_micro_stress}\\
		\mu_s&=\frac{\partial \Psi}{\partial C_s}-\nabla_R \cdot \vec{\xi}_s+\nu_s p ,\label{state_eq1_mus}\\
		\mu_i&=\frac{\partial \Psi}{\partial C_i}-\nabla_R \cdot \vec{\xi}_i + e\Phi z_i+\nu_ip,\quad i\in\mathbb{I},\label{state_eq1_mui}\\
		\vec{E}&=\frac{\partial \Psi}{\partial \vec{H}},\label{state_eq1_e}\\
		\tens{S}&=\frac{\partial \Psi}{\partial \tens{F}} - p J \tens{F}^{-T}+J(\nabla_R \cdot \vec{\xi}_J)\tens{F}^{-T}\label{state_eq1_S}.
	\end{align}\label{state_eq1}%
\end{subequations}
The energy imbalance inequality (\ref{ineq}) then reduces to:
\begin{equation}
\sum_{m\in\mathbb{M}} \nabla_R \,\mu_m \cdot \vec{J}_m \leq 0.\label{ineq_2}
\end{equation}

Having identified the thermodynamically consistent constitutive equations, in \S\ref{ent} we consider the dissipative effects associated with mass transport and determine thermodynamically consistent transport laws. 
\subsection{Derivation of transport laws} \label{ent}

Equation \eqref{ineq_2} implies that the only dissipation in the gel
arises from diffusive transport of the mobile phases down gradients of chemical potential. 
As experimental measurements to determine drag coefficients correspond to quantities in the current configuration,  we now revert to the current configuration.
The analogous form of~(\ref{ineq_2}) in the current configuration is  \cite{CIT}:
\begin{equation}
\sigma=- \sum_{m\in\mathbb{M}}c_m  \nabla \,\mu_m \cdot \bar{\vec{v}}_m \geq 0 \label{entdis2},
\end{equation}
where $\sigma$ is the rate of entropy production per unit volume of the current configuration $\mathcal{B}$, and we have exploited the relationship between the fluxes $\vec{j}_m$  and the relative velocities of the mobile species $\bar{\vec{v}}_m$ given in equation~(\ref{flux_def}). In order to establish a relationship
between the relative velocities $\bar{\vec{v}}_m$ and the gradients in the
chemical potential, we adopt a Stefan--Maxwell approach
\cite{Stefan1871,Maxwell1867} to describing multi-component diffusive
transport, which correctly captures the hydrodynamic drag (\emph{i.e.} friction) between different components of the mixture~\cite{bothe2020structure}.

To begin, the theory of linear non-equilibrium thermodynamics states that, for systems sufficiently close to equilibrium, the dissipation rate $\sigma$ is  a quadratic function of the relative velocities \cite{Rajagopal2004}:
\begin{equation}
\sigma= \sum_{m\in\mathbb{M}} \sum_{\beta\in\mathbb{M}} \ell_{m\beta} \bar{\vec{v}}_\beta \cdot \bar{\vec{v}}_m,\label{entdis3}
\end{equation}
where $\ell_{m\beta}$ are \textit{phenomenological coefficients}, chosen to satisfy the \textit{Onsager reciprocal relations} \cite{Onsager1931}. In the absence of magnetic effects, the matrix of phenomenological $\boldsymbol{\ell}=[\ell_{m\beta}]$ must be positive semi-definite.
Equations (\ref{entdis2}) and (\ref{entdis3}) then  imply that the chemical potential gradients are a linear combination of the relative velocities:
\begin{equation}
-c_m \nabla \mu_m= \sum_{\beta\in\mathbb{M}} \ell_{m\beta} \bar{\vec{v}}_\beta\label{Lcomb}.
\end{equation}
The phenomenological coefficients can be related to drag  coefficients, commonly used in mixture theory \cite{Xue2017a,Xue2017b}, by rewriting~(\ref{Lcomb}) in terms of the relative velocities between the different species:
\begin{subequations}
	\begin{align}
		-c_s \nabla \mu_s &= \sum_{i\in\mathbb{I}} f_{si} \left(\bar{\vec{v}}_s-\bar{\vec{v}}_i\right)+ f_{sn} \bar{\vec{v}}_s,\label{drag}\\
		-c_j \nabla \mu_j &= \sum_{i\in\mathbb{I}\, , i\neq j} f_{ji} \left(\bar{\vec{v}}_j-\bar{\vec{v}}_i\right) + f_{js} (\bar{\vec{v}}_j-\bar{\vec{v}}_s) + f_{jn} \bar{\vec{v}}_j,\ \ j\in\mathbb{I}\label{drag1}
	\end{align}
\end{subequations}
where $f_{ab}$ are the drag coefficients capturing the interaction between the solvent, ionic species and the polymer network. The Onsager reciprocal relations require that $f_{mb}=f_{bm}$. A common assumption in mixture theory is that the solute-solute drag can be neglected so that $f_{ij}=0$ for $i,j\in \mathbb{I}$ \cite{Xue2017a,KatzirKatchalsky1965}. The remaining drag coefficients are defined by:
\begin{equation}
f_{sn} = \frac{\nu_sc_s}{k}, \ \ f_{js}=\frac{k_BT c_j}{\D^0_{j}},\ \  f_{js}+f_{jn}= \frac{k_BT c_j}{\D_j}, \label{drag2}
\end{equation}
where $k$ is related to the permeability of the solvent in the network, $\D^0_j$ is the diffusion coefficient of the solute in pure solution, while $\D_j$ is the diffusion coefficient of the solute in the gel. For this choice of drag coefficients, the matrix $\boldsymbol{\ell}$ is:
\begin{equation}
\boldsymbol{\ell}=\begin{bmatrix}
\mathbf{d} & -\vec{f}_s\\
-\vec{f}_s^T & d_s
\end{bmatrix},
\quad \vec{f}_s=\begin{bmatrix}
f_{1s}\\
\vdots\\
f_{Ns}
\end{bmatrix}
, \quad d_s=\sum\limits_{i} f_{si}+f_{sn}\label{lmatrix}
\end{equation}
where $\mathbf{\ell}$ is a symmetric diagonally dominant matrix with positive diagonal entries and hence positive semi-definite in line with the Onsager reciprocal relations. 

Using~(\ref{drag})-(\ref{drag2}), together with (\ref{flux_def}) the expressions for the fluxes  $\vec{j}_m$ in terms of the chemical potential gradients are:
\begin{subequations}
	\begin{align}
		{\vec{j}}_s &=c_s\bar{\vec{v}}_s=-\frac{c_s K}{\nu_s}  \left(\nabla \mu_s +\sum_{i\in\mathbb{I}} \frac{\D_i}{\D^0_i} \frac{c_i}{c_s} \nabla \mu_i\right),\label{vbar2a}\\
		{\vec{j}}_i &= c_i\bar{\vec{v}}_i= {-}\frac{\D_ic_i}{k_B T}\nabla \mu_i {+} \frac{\D_i c_i}{\D^0_i c_s} \vec{j}_s\, \ \ i\in\mathbb{I}\label{vbara},
	\end{align}
	where the coefficient $K$  is defined to be
	\begin{align}
		\frac{1}{K} = \frac{1}{k} + \sum_{i\in\mathbb{I}} \frac{k_B T}{\nu_s\D^0_i} \left(1-\frac{\D_i}{\D^0_i}\right) \frac{c_i}{c_s}.\label{eq:K}
	\end{align}
\end{subequations}
Here $K$ represents the Darcy hydraulic permeability (over dynamic viscosity) of the gel to the solvent and ionic species, whilst $k$ represents the Darcy hydraulic permeability (over dynamic viscosity) to pure solvent.
\subsection{The full polyelectrolyte gel model}\label{gelcurrent}

We now present the full polyelectrolyte gel model in the current configuration. The system of governing equations in the reference configuration is given in \ref{eqref}. We start by using the specific form of the free energy to determine the constitutive equations, and we discuss their physical interpretation in \S\ref{state1}. In \S\ref{full} we then collect these equations together with the conservation of mass and momentum equations, definition of electric potential and Gauss' law of electrostatics, and summarise the full gel model.

\subsubsection{Constitutive equations for prescribed free energy}\label{state1}
Substituting the precise form of the free energy $\Psi$, given in \S\ref{free_energy}, into equations (\ref{state_eq1}), we obtain the constitutive equations in the reference configuration (see \ref{eqref}).  Their form simplifies when expressed in terms of the current configuration.  

The expressions for the microstresses (equations~(\ref{state_eq1_micro_stress})) are given by 
\begin{subequations}
	\begin{align}
		\boldsymbol{\xi}_s =  \tens{F}^{-1}\nabla\,c_s,\label{microstress1} \\[2mm]
		\boldsymbol{\xi}_J =-c_s\tens{F}^{-1}\nabla\,c_s,\label{microstress2}
	\end{align}
\end{subequations}
These expressions will be utilised in determining the expressions for the solvent chemical potential $\mu_s$ and the Cauchy stress $\tens{T}$ in the analysis that follows. 

The chemical potential for the solvent (equation~(\ref{state_eq1_mus})) is given by
\begin{subequations}
	\begin{align}
		\mu_s=\mu_s^0+\nu_s(p+\Pi_s)+\mu^G_s,\label{solventchempot}\end{align}
	where the osmotic pressure $\Pi_s$ and $\mu_s^G$ are the contributions to
	arising from the mixing energy and solvent concentration gradient, respectively, and they are defined as
	\begin{align}
		\Pi_s=\frac{k_BT}{\nu_s}\left[\frac{\chi(1-\nu_s c_s)}{J}+\ln (\nu_s c_s) +1 -\sum_{m\in\mathbb{M}} \nu_sc_m\right], \label{mus}\\
		\mu_s^G= - \gamma\nabla^2c_s, \label{gradnew}
	\end{align}
\end{subequations}
where $J$ is given by~(\ref{incompcurrent}). Similarly, the electrochemical potential for the ionic species (equation (\ref{state_eq1_mui})) is given by
\begin{subequations}
	\begin{align}
		\mu_i=\mu_i^0+\nu_i(p+\Pi_i)+z_i e\Phi,\ i\in\mathbb{I}\label{ionschempot}\end{align}
	where the osmotic pressure $\Pi_i$ is defined by 
	\begin{align}
		\Pi_i=  \frac{k_BT}{\nu_i} \left[ -\frac{\chi c_s\nu_i}{J} +\ln (\nu_ic_i)+1-\sum_{m\in\mathbb{M}}\nu_ic_m\right], \ i\in\mathbb{I},\label{mui}
	\end{align}
	and characterises the contributions to the electrochemical potential arising from the mixing energy. 
\end{subequations}

Having the form of the solvent chemical potential (\ref{mus}) and ionic species electrochemical potential (\ref{mui}) then enables us to determine the solvent and ionic species fluxes, given by (\ref{vbar2a}) and (\ref{vbara}). The dependencies of the (electro)chemical potentials on the mobile species concentrations, the gradient of the solvent concentration, the thermodynamic pressure and the electrostatic potential, as well as the deformation of the gel,  means that the solvent and ionic species fluxes are complex expressions. 
However, as we have shown for a neutral hydrogel \cite{Hennessy2020}, it is straightforward  to reduce these fluxes to familiar forms in simplifying limits. Neglecting for now terms due to interfacial energies and the electric field, we can consider two sublimits. 

Firstly, we consider $\nu_s C_s\rightarrow\infty$, which corresponds to the limit of a large number of solvent molecules. In the current configuration this corresponds to $\nu_s c_s=\phi_s\rightarrow 1$, and $\nu_i c_i=\phi_i\rightarrow 0$ where $\phi_m, \ m\in\mathbb{M}$, are the volume fractions of the mobile species in the current configuration.
Retaining leading-order terms, we find that
\begin{equation}
\vec{j}_s=-c_sk\nabla p.
\end{equation}
Darcy's law is usually written in the form 
$\vec{v}_s-\vec{v}_n=-k\nabla p$.
The latter has be experimentally estimated (see e.g.~\cite{Drozdov2016b} and discussion in \cite{Bertrand2016,Hennessy2020}) and it is found to have the form:
\begin{equation}
k(J)= \frac{\D_s\nu_s}{k_BT}J^{\theta}\label{k_def}.
\end{equation}
where $\theta$ is a positive constant.

Secondly, we can consider the limit of a small number of molecules of mobile species, $\nu_m C_m\rightarrow 0, \ m\in\mathbb{M}$, so that in the current configuration $\phi_m\rightarrow 0,\ m\in\mathbb{M}$, and $J\rightarrow 1$. Retaining only the leading-order terms as $\phi_m\rightarrow 0$, and given that, based on~(\ref{k_def}), $k\rightarrow \D_s\nu_s/(k_BT)$, we find that
\begin{equation}
\vec{j}_s=- \D_s\left(\nabla c_s+\sum\limits_{i\in\mathbb{I}}\frac{\mathcal{D}_i}{\mathcal{D}_i^0}\nabla c_i\right),
\end{equation} 
so that we recover Fick's law of diffusion (first term) and osmotic flux of solvent due to gradients in ionic concentrations  (second term). Similar expressions can be found for the flux of ionic species in this limit. 

From~(\ref{state_eq1_e}), we find 
\begin{equation}
\vec{e} = \frac{1}{\epsilon }\vec{h}\, ,\label{relel}\\
\end{equation}
which captures the standard relationship between electric field and displacement for a linear, homogeneous, isotropic dielectric with instantaneous response to changes in the electric field. 

Finally, the Cauchy stress tensor is expressed as follows
\begin{subequations}
	\begin{align}
		\tens{T}= -p \tens{I} + \tens{T}_K + \tens{T}_M+  \tens{T}_e ,\label{Cauchy} \\
		\tens{T}_K=\gamma\left[\left(\frac{|\nabla c_s|^2}{2}+c_s\nabla^2c_s\right)\tens{I} - \nabla c_s \otimes \nabla c_s\right], \label{Korteweg}\\
		\tens{T}_M=\epsilon \left[\nabla \Phi \otimes \nabla \Phi-\frac{1}{2} \,|\nabla \Phi|^2\tens{I} \right], \label{Maxwell}\\
		\tens{T}_e = \frac{G}{J}\left(\tens{B} - \tens{I}\right).
	\end{align}\label{eq:Stressgel}%
\end{subequations}
In  determining the Cauchy stress $\tens{T}$ in (\ref{Cauchy}) we used the expression for the microstress $\boldsymbol{\xi}_J$ given by (\ref{microstress2}). The first term in (\ref{Cauchy}) is the isotropic stress induced by the pressure, while the second term is the elastic stress of the network. The Korteweg  (\ref{Korteweg}) $\tens{T}_{K}$ and  Maxwell  (\ref{Maxwell})  $\tens{T}_M$ stress tensors capture the stresses generated within the gel due to the formation of  internal interfaces (i.e., gradients of the solvent concentration) and the presence of the electric field, respectively.

\subsubsection{Full model summary}\label{full}

For ease of reference, we present now the complete model in the current configuration. 
The conservation equations in the current configuration are given in Section \ref{conslaws} by equations (\ref{incompcurrent}), (\ref{consmass_current}), (\ref{consmomcurrent}), (\ref{phicurrent}) and (\ref{gausscurrent}).
We use (\ref{relel}) to combine equations (\ref{phicurrent})  and (\ref{gausscurrent}) via the elimination of the electric field.
The full polyelectrolyte gel model in the current configuration is then given by 
\begin{subequations}\label{fullfinal}
	\begin{align}
		J=\left(1-\sum_m \nu_m c_m\right)^{-1},& \\
		\partial_t c_s + \nabla \cdot(c_s \vec{v}_n)=-\nabla\cdot\vec{j}_s,&\\
		\partial_t c_i + \nabla \cdot(c_i \vec{v}_n)= -\nabla\cdot\vec{j}_i,&\ \ i\in \mathbb{I}\label{consmassfinal}\\
		\nabla \cdot \tens{T}=\mathbf{0},&\\
		-\epsilon\nabla^2 \Phi= e\left(\sum\limits_{i\in\mathbb{I}} z_i c_i+z_f c_{f}\right),&\label{Poissongel}
	\end{align}
	where the diffusive fluxes are
	\begin{align}
		{\vec{j}}_s =-c_s K(c_s,c_i,J)  \left(\nabla \mu_s +\sum_{i\in\mathbb{I}} \frac{\D_i}{\D^0_i} \frac{c_i}{c_s} \nabla \mu_i\right),&\\
		{\vec{j}}_i = -\frac{\D_i c_i}{k_B T}\nabla \mu_i + \frac{\D_i c_i}{\D^0_i c_s} {\vec{j}}_s,&\ \ i\in\mathbb{I} 
	\end{align}
	together with the constitutive equations
	\begin{align}
		\mu_s = \mu_s^0 + \nu_s (p+\Pi_s) - \gamma\nabla^2c_s,& \label{order}\\
		\mu_i = \mu^0_i +\nu_i (p+\Pi_i) + z_i e \Phi,& \quad i\in\mathbb{I}\\
		\Pi_s=\frac{k_BT}{\nu_s}\left[\frac{\chi(1-\nu_s c_s)}{J}+\ln (\nu_s c_s) +1 -\sum_{m\in\mathbb{M}} \nu_sc_m\right],& \\
		\Pi_i=  \frac{k_BT}{\nu_i} \left[ -\frac{\chi c_s\nu_i}{J} +\ln (\nu_ic_i)+1-\sum_{m\in\mathbb{M}}\nu_ic_m\right], &\quad i\in\mathbb{I},\\
		\tens{T}= -p \tens{I} +  \tens{T}_K +\tens{T}_M + \tens{T}_e,\\
		\tens{T}_K=\gamma\left[\left(\frac{|\nabla c_s|^2}{2}+c_s\nabla^2c_s\right)\tens{I} - \nabla c_s \otimes \nabla c_s\right], &\\
		\tens{T}_M=\epsilon \left[\nabla \Phi \otimes \nabla \Phi-\frac{1}{2} \,|\nabla \Phi|^2\tens{I} \right],& \\
		\tens{T}_e = \frac{G}{J}\left(\tens{B} - \tens{I}\right).&
	\end{align} \label{full_mod_gel}%
\end{subequations}
The network velocity satisfies $\vec{v}_n=\partial\vec{u}/\partial t +(\vec{v}_n\cdot\nabla)\vec{u}$ and the Darcy hydraulic permeability $K$ is defined by Equations~(\ref{eq:K}) and (\ref{k_def}).
Furthermore, $\tens{B} = \tens{F}\tens{F}^T$ where $\tens{F} = (\tens{I} - \nabla \vec{u})^{-1}$.

\section{Model derivation for an ionic bath}

Having derived the model for the polyelectrolyte gel, we now  derive the governing equations for the ionic bath adjacent to the gel, following the approach we adopted for the gel.  The bath is considered as a two-phase isotropic mixture with solvent and solute molecules, where the latter are considered to be ionic charges. As the bath has no solid component, we do not have a natural reference configuration. For this reason, the model is derived directly in the current configuration, as standard in mixture theory \cite{bowen1976theory}. 
The notation for variables in the current configuration that are common to both the polyelectrolyte gel and ionic bath, {\it e.g.}~the ionic concentrations, remains the same. 
In \S\ref{bath_cons_law} we present the general model set up and state the conservation laws for each component of the mixture, as well as the mixture as a whole. The constitutive properties of the system are specified in \S\ref{bath_free_energy} via the definition of the Helmholtz free energy. The governing and constitutive equations of the system are then derived by using thermodynamics arguments in \S\ref{bath_en_imb}. We  conclude with a summary of the full model in \S\ref{fullbath}, and discuss how our model links to previously proposed models.

\subsection{Conservation equations and electrostatics in the current configuration}
\label{bath_cons_law}
As shown in Figure \ref{gel_schem}, the ionic bath consists of the solvent and the $N$ freely moving ionic species. Following the same convention as for the gel derivation, we use the subscript $s$ to denote the solvent, and the index $i=1, \dots, N$ to denote the ionic species. As in Section~\ref{ModelSetUp}, the set $\M$ is given by the collection of indices $\M=\left\{s,1,\ldots,N\right\}$.

Assuming each species in the mixture to be incompressible, the transport law for component reads:
\begin{subequations}
	\begin{align}
		\frac{\partial c_m}{\partial t} + \nabla \cdot \left(c_m \vec{v}_m\right)=0, \quad m\in\M. \label{mass_cons1}
	\end{align} 
	Moreover, we assume that each point in the mixture is occupied by solvent and/or solute, \emph{i.e.} no voids can form in the mixture, resulting in the no-void condition: 
	\begin{align}
		1=\sum_{m\in\M} \nu_m c_m.\label{no_void}
	\end{align}
	We decompose the total flux of the $m$-th species into an advective component that is carried with the mean velocity of the mixture and a diffusive component which represents transport down gradients in chemical potential.
	The mean mixture velocity is defined as
	\begin{align}
		\vec{v}=\sum_{m\in \M} \nu_m c_m \vec{v}_m,\label{vel_mix}
	\end{align}
	and the flux of the $m$-th species relative to the mixture velocity is defined as $\vec{q}_m=c_m\left(\vec{v}_m-\vec{v}\right)$.
	Equation (\ref{mass_cons1}) is then:
	\begin{align}
		\frac{\partial c_m}{\partial t} + \nabla \cdot \left(c_m \vec{v}\right)=-\nabla \cdot \vec{q}_m, \ m\in\M.\label{mass_cons2}
	\end{align}
	Using the definition of mixture velocity~(\ref{vel_mix}), together with the no-void condition (\ref{no_void}) we have that the fluxes must satisfy:
	\begin{align}
		\sum_{m\in \M} \nu_m\vec{q}_m=\vec{0}.\label{flux_sum}
	\end{align}
	Thus in place of $\vec{v}_m$ we now use $\vec{v}$ and $\vec{q}_m$, $m\in\M$. 
	We introduce the material derivative $D(\cdot)/Dt= \partial(\cdot)/\partial t + (\vec{v}\cdot\nabla)(\cdot)$, so that
	equation~(\ref{mass_cons2}) takes the simple form:
	\begin{align}
		\frac{D c_m}{Dt}+c_m \nabla \cdot \vec{v} =-\nabla \cdot \vec{q}_m, \ m\in\M.
	\end{align}
	By multiplying \eqref{mass_cons2} by $\nu_m$, summing over $m$, and using
	\eqref{no_void} and \eqref{flux_sum}, we find that the mixture velocity is divergence free:
	\begin{align}
		\nabla \cdot \vec{v} = 0. \label{eqn:ic}
	\end{align}
\end{subequations}

Neglecting the inertia of the mixture (so that viscous effects dominate inertial effects corresponding to low Reynolds number of the mixture), and assuming that the bath is not subject to external forces,  conservation of momentum for the mixture is:
\begin{equation}
\nabla \cdot \tens{T}=\vec{0},
\end{equation}
where $\tens{T}$ is the stress tensor. Following \cite{Coleman1963, Groot1962}, we assume the stress tensor is symmetric, which implies balance of internal and external angular momentum; for a more detailed discussion, we refer to \cite{Groot1962}. 

Finally, the relationship between the electric field $\vec{e}$  and electrostatic potential $\Phi$, and Gauss' law of electrostatics for the electric displacement  $\vec{h}$ are analogous to those  for the gel~(\ref{phicurrent})-(\ref{gausscurrent}) with the charge density $q$ given by:
\begin{equation}
q=\sum_{i\in\I}e z_i c_i.
\end{equation}

To facilitate the derivation of state and constitutive equations for the bath,
we introduce the velocity gradient tensor $\tens{L}$, which is defined as $\tens{L}=\nabla \vec{v}$. Using this notation, we can write $\nabla \cdot \vec{v}=\tens{I}:\tens{L}=\mbox{tr}(\tens{L})$, where $\mbox{tr}(\cdot)$ denotes the trace of a tensor. 
The tensor $\tens{L}$ is commonly decomposed as the sum of its symmetric ($\tens{D}$) and skew symmetric ($\tens{W}$) parts:
\begin{equation}
\tens{D}=\frac{\tens{L}+\tens{L}^T}{2},\quad \tens{W}=\frac{\tens{L}-\tens{L}^T}{2}.
\end{equation}
While the tensor $\tens{D}$ is commonly called the \textit{rate of deformation}, $\tens{W}$ is known as the \textit{vorticity} or \textit{spin} tensor.
\subsection{Free energy}
\label{bath_free_energy}
The Helmholtz free energy per unit volume of the mixture (note that in \S\ref{free_energy} the free energies were per unit volume in the reference configuration) is denoted by $\psi=\psi(\vec{x},t)$, and has three contributions:
\begin{equation}
\psi=\psi_1+\psi_2+\psi_3,
\end{equation}
corresponding to the energy of the electric field ($\psi_1$); the energy of solvent and solutes not interacting with each other ($\psi_2$); and the energy associated with mixing the different component of the mixture ($\psi_3$). Note that we have here neglected the interfacial energy  as we do not anticipate phase separation occurring in the ionic bath. Having discussed the forms of these free energies in detail in \S\ref{free_energy}, here we give the specific forms of each contribution. The first energy  $\psi_1$ is 
\begin{subequations}
	\begin{align}
		\psi_1=\frac{1}{2\epsilon} \vec{h}\cdot \vec{h},
	\end{align}
	where $\epsilon$ is the permittivity of the mixture. As for the polyelectrolyte gel, the permittivity is dominated by that of the solvent, with little contribution from the mobile phases.   For this reason, we consider the permittivity of the both the polyelectrolyte gel and ionic bath to be equal.  
	
	The energy density $\psi_2$ is given by:
	\begin{align}
		\psi_2 = \sum\limits_{m\in \M} \mu^0_m c_m,
	\end{align} 
	with $\mu^0_m$ as defined in \S\ref{free_energy}. Finally the mixing energy of the solvent and ions $\psi_3$ reads:
	\begin{align}
		\psi_3 = k_B T \sum\limits_{m\in \M}c_m \ln\left( c_m\nu_m\right).
	\end{align} 
\end{subequations}

\subsection{Energy imbalance inequality}
\label{bath_en_imb}
Considering a control volume in the current configuration $\mathcal{V}(t)$, the energy imbalance inequality~(\ref{energyin}) is now:
\begin{equation}
\frac{\d}{\d t} \left\{\int_{\mathcal{V}(t)} \psi \, dv\right\}\leq \mathcal{W}(\mathcal{V}(t)) + \mathcal{M}(\mathcal{V}(t)). \label{curr_ibm}
\end{equation}

This inequality must hold for all motions which
satisfy the no-void \eqref{no_void} and incompressibility \eqref{eqn:ic}
constraints.
However, since these two constraints are algebraically equivalent, it is
sufficient to impose only one of them when using the energy imbalance inequality
to derive constitutive relationships for the bath. We choose to
enforce the no-void condition \eqref{no_void}
via a Lagrange multiplier. This approach
allows the composition variables $c_m$ to be treated as independent and does
not require the incompressibility constraint
$\nabla \cdot \vec{v} = \tens{I}:\tens{L} = 0$ to be enforced during the
calculations. Alternative
derivations which do enforce the incompressibility constraint,
rather than the no-void constraint, can be found in \cite{Kim2005}.
Thus, by using Reynolds' transport theorem, the energy imbalance inequality \eqref{curr_ibm} can be written as:
\begin{equation}
\int_{\mathcal{V}(t)} \left[\frac{D\psi}{Dt} +\psi\left(\tens{I}:\tens{L}\right)\right]dv\leq \mathcal{W}(\mathcal{V}(t)) + \mathcal{M}(\mathcal{V}(t)). \label{bath_en_imb1}
\end{equation}
The rate of mass transport $\mathcal{M}$ is analogous to that for the gel, {\it i.e.}~Equation~(\ref{mass-rate}), but expressed in the current configuration. 
We therefore obtain:
\begin{subequations} \label{rates-work+mass}
	\begin{align}
		\mathcal{M}(\mathcal{V}(t))
		&=\sum\limits_{m\in \M} - \int_{\mathcal{V}(t)} \nabla \cdot\left( \mu_m \,\vec{q}_m\right) dv \label{bath-mass-rate}
	\end{align}
	where $\mu_m$ is the chemical potential associated with each species in the solution. Similarly, we can derive the rate of electrical and mechanical from~(\ref{electricalwork})-(\ref{mechanicalwork}) respectively by moving to the current state
	to find
	\begin{align}
		\hspace{-5mm}\mathcal{W}_{el}(\mathcal{V}(t)) &=
		\int_{\mathcal{V}(t)} \left(\vec{e}\cdot\frac{D\vec{h}}{Dt}+\left\{\left[\vec{e}\cdot\vec{h}-\Phi(\nabla\cdot \vec{h})\right]\tens{I}-\vec{e}\otimes \vec{h}\right\}:\tens{L}-\Phi \frac{Dq}{Dt}\right) dv,
		\label{electricalionic}\\
		\mathcal{W}_{mec}(\mathcal{V}(t)) &
		=\int_{\mathcal{V}(t)}  (\tens{T}:\tens{L}) dv. \label{mechanicalionic}
	\end{align}
\end{subequations}
The derivation of (\ref{electricalionic}) and (\ref{mechanicalionic}) can be
found in \ref{AppIonic}.

From \S\ref{bath_free_energy}, the free energy $\psi$ takes the form $\psi=\psi(c_m,\vec{h})$. Therefore, given the definitions~(\ref{rates-work+mass}) and following the same steps as in Section~\ref{sec_ine}, we rewrite~(\ref{bath_en_imb1}) as:
\begin{subequations}
	\begin{align}
		\begin{aligned}
			\left(\tens{T}_{equi}-\tens{T}\right):\tens{L}+\left(\frac{\partial\psi}{\partial\vec{h}} -\vec{e}\right)\cdot\frac{D\vec{h}}{Dt}+\sum_{i\in\I}\left[\Phi e z_i-\mu_i +\nu_i\lambda+\frac{\partial\psi}{\partial c_i}\right] \frac{Dc_i}{Dt}\\
			+\left(\nu_s\lambda-\mu_s+\frac{\partial\psi}{\partial c_s}\right)\frac{Dc_s}{Dt}
			+\sum_{m\in\M} \nabla \mu_m \cdot \vec{q}_m \leq 0,
		\end{aligned}\label{longnow}
	\end{align}
	where we have introduced the Lagrange multiplier $\lambda$ to account for the no-void condition~\eqref{no_void} and we define $\tens{T}_{equi}$ as
	\begin{align}
		\tens{T}_{equi}=-(\vec{e}\cdot\vec{h})\tens{I}+\Phi(\nabla\cdot \vec{h})\tens{I}+\vec{e}\otimes \vec{h}+\left(\psi-\sum_{m\in\M}  \mu_mc_m\right)\tens{I}.\label{eq:Teq}
	\end{align}
\end{subequations}
In the gel, the viscosity of the fluid is captured in the permeability term, {\it i.e.}~via dissipation due to the motion of the solvent with respect to the gel. In the bath, this needs to be accounted for explicitly and we therefore decompose the stress tensor as the sum of an equilibrium $\tens{T}_{equi}$ (as defined in Eq.~\ref{eq:Teq}) and viscous contribution $\tens{T}_v$, \emph{i.e.} $\tens{T}=\tens{T}_{equi}+\tens{T}_v$.

Similarly to the gel, inequality \eqref{longnow} is linear in $D\vec{h}/Dt$, $Dc_m/Dt$, $m\in\M$, each of which can be chosen independently at each point $\vec{x}$ and time $t$.  For the energy imbalance inequality to be always satisfied, we must have that:
\begin{subequations}
	\begin{align}
		\mu_s &= \frac{\partial \psi}{\partial c_s}+\nu_s\lambda,\\
		\mu_i &= \frac{\partial \psi}{\partial c_i} +e\Phi z_i +\nu_i\lambda,\ i\in\I\\
		\vec{e}&=\frac{\partial \psi}{\partial \vec{h}}.
	\end{align}\label{bath_const_1}
\end{subequations}
The energy imbalance (\ref{longnow}) then reduces to:
\begin{equation}
-\tens{T}_v : \tens{L}+ \sum_{m\in \M}\nabla\,\mu_m\cdot \vec{q}_m \leq 0 .\label{in_bath1}
\end{equation}
In the following section, we consider the irreversible processes associated with mass transport and viscous dissipation, and determine the equations relating the mass fluxes $\vec{q}_m$ to the chemical potential gradients $\nabla\mu_m$, and the viscous stress tensor $\tens{T}_v$ to the velocity gradient tensor $\tens{L}$. 

\subsubsection{Derivation of transport laws}

Given the symmetry of the stress tensor $\tens{T}$ and the fact that $\tens{T}_{equi}$ is symmetric (as verified below), we must have that $\tens{T}_v$ is symmetric. 
We then have that $\tens{T}:\tens{W}=0$, so that $\tens{L}=\tens{D}+\tens{W}$ can be substituted by $\tens{D}$ in~(\ref{in_bath1}). Utilising the definition of the mixture velocity, {\it i.e.}~equation~(\ref{vel_mix}), together with the relationship $\vec{q}_m=c_m(\vec{v}_m-\vec{v})$, equation (\ref{in_bath1}) can be rewritten as:
\begin{equation}
-\tens{T}_v :  \tens{D}+ \sum\limits_{m\in \M} c_m\left(\nabla\,\mu_m- \sum_{\beta\in \M} c_\beta \nu_m \nabla\,\mu_\beta\right)\cdot \vec{v}_m \leq 0.
\end{equation}
Again assuming that we are in the regime of linear non-equilibrium thermodynamics (see \S\ref{ent}), and considering the additional constraint imposed by \textit{Curie's law}\footnote{Macroscopic causes cannot have more elements of symmetry than the effect they cause (i.e. there cannot be any coupling
	between thermodynamic variables of different tensorial nature)}, we arrive at the following set of force-flux relations:
\begin{subequations}
	\begin{align}
		\tens{T}_v=2\eta \left(\tens{D} -\frac{1}{3}(\tens{I}:\tens{D})\tens{I}\right) + \kappa(\tens{I}:\tens{D})\tens{I},& \label{T_v_full}\\
		-c_m\left(\nabla\mu_m- \sum_{\beta\in\M} \nu_mc_\beta \nabla\mu_\beta\right) = \sum_{k\in\M} \ell_{km} \vec{v}_k,&\quad m\in\M\label{friction}
	\end{align}
\end{subequations}
where the matrix of phenomenological coefficients $\boldsymbol{\ell}=[\ell_{ij}]$ must again be symmetric and semi-positive definite, while $\eta$ and $\kappa$ are positive constants representing the shear and dilatational viscosity of the bath, respectively. Note that the latter will not actually play a role as the isotropic component of $\tens{T}_v$ will vanish upon strongly imposing the incompressibility condition~(\ref{eqn:ic}).
As before (see \S \ref{ent}), we can recover Stefan--Maxwell diffusion \cite{Stefan1871,Maxwell1867} by relating the phenomenological coefficients $\ell_{ij}$ to the drag coefficients commonly used in mixture theory. To do so, we rewrite equation (\ref{friction}) in terms of the relative velocities between the different phases: 
\begin{equation}
-c_m\left(\nabla\mu_m- \sum_{\beta\in \M} \nu_mc_\beta \nabla\mu_\beta\right) = \sum_{k\in\M\setminus\{m\}} f_{km} \left(\vec{v}_m-\vec{v}_k\right), m\in\M.\label{flux_mix}
\end{equation} 
We again consider the drag between ions to be negligible, while $f_{si}=f_{is}$ are defined as in~(\ref{drag2}).  Analogously to the result in \S \ref{ent}, for this choice of the fluxes, we have that the matrix $\boldsymbol{\ell}$ has the same structure as in Eq.~(\ref{lmatrix}) by setting $f_{mn}=0$ for all $m\in \M$. Therefore $\boldsymbol{\ell}$ is still a symmetric and diagonally dominated matrix and hence positive semi-definite.

The system~(\ref{flux_mix}) is, in fact, under-determined (their sum is indeed identically zero). This is to be expected as the velocities are not independent of each other but need to satisfy~(\ref{flux_sum}). We therefore use~(\ref{flux_mix}) to express the ionic velocities in terms of $\vec{v}_s$, where the latter is defined by~(\ref{flux_sum}). All this considered, we obtain:
\begin{equation}
\vec{v}_i = -\frac{\D^0_i}{k_BT}\left(\nabla \mu_i -\sum_{\beta\in \M} \nu_ic_\beta\nabla \mu_\beta\right) + \vec{v}_s, \ i\in\I.
\end{equation}
Simply subtracting the mixture velocity and multiplying by the ionic concentration $c_i$ we get:
\begin{subequations}
	\begin{align}
		\vec{q}_i& = -\frac{\D^0_ic_i}{k_BT}\left(\nabla \mu_i -\sum_{\beta\in \M} \nu_ic_\beta\nabla \mu_\beta\right) + \frac{c_i}{c_s} \vec{q}_s,\\
		\vec{q}_s &=-\sum_{i\in I} \frac{\nu_i}{\nu_s}\vec{q}_i,\label{qs}
	\end{align} \label{fluxq}
\end{subequations}
where~(\ref{qs}) is obtained by simply readjusting the terms in~(\ref{flux_sum}). 

\subsubsection{Constitutive Equations}
Using the definition of the Helmholtz free energy given in \S\ref{bath_free_energy}, the constitutive laws~(\ref{bath_const_1}) are as follows:
\begin{subequations}
	\begin{align}
		\mu_s =\mu_s^0 + \nu_s(p+\Pi_s),\\ \quad \mu_i =\mu^0_i+\nu_i (p+\Pi_i) +  z_i e\Phi,&\quad i\in\I,\label{chem1} \\
		\Pi_m= \frac{k_BT}{\nu_m} \left[\ln\left(\nu_m c_m\right)+1-\sum_{j\in\M} \nu_m c_j\right],&\quad m\in\M.\end{align}
	As before, $\mu_s$ is the chemical potential for the solvent and has contributions arising from the thermodynamic pressure $p$ (to be defined below) and the osmotic pressure $\Pi_s$. Similarly, $\mu_i$ is the electrochemical potential for the ionic species, with the thermodynamic and osmotic pressures, and the electrostatic potential all contributing. 
	The relationship between the electric field and displacement is again given by
	$\vec{h}=\epsilon\vec{e}$.
	
	The symmetric tensor $\tens{T}_{equi}$  is found to be
	\begin{align}
		\tens{T}_{equi}=-p\tens{I}+\tens{T}_M,\end{align}
	where
	\begin{align}
		\tens{T}_M=\epsilon\left[\nabla\Phi\otimes\nabla\Phi-\frac{|\nabla \Phi|^2}{2}\tens{I}\right],
	\end{align}
	is the Maxwell stress tensor (see \S\ref{state1}) due to the presence of the electric field, and the thermodynamic pressure $p$ is defined as
	$p=\lambda+ k_BT\sum_{m\in \M} c_m$.
\end{subequations}

\subsection{The full ionic bath model \label{fullbath}}
The complete model for the bath is given by the following conservation laws presented in Section~\ref{bath_cons_law}:
\begin{subequations}
	\begin{align}
		\partial_t c_i+\nabla\cdot \left(c_i \vec{v}\right)=-\nabla \cdot \vec{q}_i,& \quad i\in \I,\\
		\nabla \cdot \tens{T}=\vec{0},&\label{cons_mom_bath}\\
		-\epsilon \nabla^2 \Phi = e \sum_{i\in \I} z_ic_i,&\label{poissonbath}
	\end{align}
	where the fluxes are defined by:
	\begin{align}
		\vec{q}_i = -\frac{\D^0_ic_i}{k_BT}\left(\nabla \mu_i -\sum_{\beta\in \M} \nu_ic_\beta\nabla \mu_\beta\right) + \frac{c_i}{c_s} \vec{q}_s,&\quad i\in\mathbb{I}\\
		\vec{q}_s =-\sum_{i\in \I} \frac{\nu_i\vec{q}_i}{\nu_s},&
	\end{align}
	and the mixture velocity $\vec{v}$ satisfies the standard incompressibility condition:
	\begin{align}
		\nabla \cdot \vec{v}=0.
		\label{ic_bath}
	\end{align}
	Note that we do not need to account for the governing equation for the solvent 
	concentration $c_s$, as this is computed using the no-void condition~(\ref{no_void}):
	\begin{align}
		c_s=\frac{1-\sum_{i\in \I} \nu_i c_i}{\nu_s}.
	\end{align}
	The model is then completed by specifying the constitutive equations:
	\begin{align}
		\mu_s = \mu_s^0 +\nu_s p+ k_BT\left[\ln (\nu_s c_s)+1-\sum_{m\in\M} \nu_s c_m\right],&\label{chem1new}\\
		\mu_i = \mu^0_i+\nu_i p+ z_i e\Phi + k_BT \left[\ln (\nu_ic_i)+1-\sum_{m\in\M} \nu_ic_m\right],&\quad i\in\I,\label{chem2new}\\
		\tens{T}=-p\tens{I}+\tens{T}_M+\tens{T}_v,&\label{tens1_bath}\\
		\tens{T}_M=\epsilon\left[\nabla\Phi\otimes\nabla\Phi-\frac{|\nabla \Phi|^2}{2}\tens{I}\right],&\label{tens2_bath}\\
		\tens{T}_v=\eta \left(\nabla \vec{v}+\nabla \vec{v}^T\right),&\label{tens3_bath}
	\end{align}\label{bath_model}%
	and initial conditions for the concentration of ions $c_i$.
	The viscous stress tensor \eqref{tens3_bath} has been simplified from
	\eqref{T_v_full} by making use of the tensorial form of the
	incompressibility condition
	\eqref{ic_bath}, which reads $\tens{I}:\tens{D} = 0$.
\end{subequations}

To better understand the phenomena that are driving the flow in the bath, we can rearrange the governing equations to show how the flux of the ionic species depends on gradients of the ionic concentration and electrostatic potential.  We  use~(\ref{cons_mom_bath}) and (\ref{tens1_bath})-(\ref{tens3_bath}) to obtain the following expression for the pressure gradient:
\begin{equation}
\nabla p= \eta \nabla^2 \vec{v} + \epsilon \nabla \Phi \nabla^2\Phi= \eta \nabla^2 \vec{v}- \left(\sum_{i\in \I} e c_i z_i\right) \nabla \Phi.\label{simplify}
\end{equation}
Equations~(\ref{chem1new})-(\ref{chem2new}) can then be used to derive the following relationship between the gradients of the  chemical potentials:
\begin{equation}
\sum\limits_{m\in \M}c_m\nabla \mu_m=\eta \nabla^2 \vec{v}.\label{bal}
\end{equation}
If we consider the case in which all of the molecular volumes $\nu_m$ are equal to $\nu$ and the well-known electro-neutral limit, i.e. $\sum_{i\in\I} z_ic_i=0$, then, using (\ref{simplify}) and (\ref{bal}), we can reduce the fluxes $\vec{q}_i $ to
\begin{equation}
\vec{q}_i = -\D^0_i\left[\nabla c_i+\frac{ez_ic_i}{k_BT} \nabla \Phi\right] + \frac{c_i}{c_s} \vec{q}_s,\quad i\in\I.
\end{equation}
This is the classic Nernst--Planck equation with an additional advection contribution due to cross-diffusion of the mixture components as captured by our Stefan--Maxwell approach.

\section{Interfacial conditions}\label{IntCond}
The behaviour of the polyelectrolyte gel and ionic bath domains are coupled via specification of interfacial boundary conditions.  
We denote the position of the interface {\it in the current configuration} by $\Gamma$, while $[\cdot]^{+}_-$ denotes the jump in the value of a variable across the interface where $-$ and $+$ stand for the limit approaching from the gel and the bath domain respectively. The local velocity of the interface
$\vec{v}_\Gamma$ is equivalent to the normal component of the network velocity
$\vec{v}_n$. Thus the kinematic boundary condition reads:
\begin{subequations}
	\begin{equation}
	\vec{v}_\Gamma = \left(\vec{v}_n \cdot \vec{n}\right) \vec{n},
	\end{equation}
	where $\vec{n}=\vec{n}(\vec{x},t)$ is the normal vector to the interface. 
	Consequently, imposing the conservation of mass across the interface and using a pillbox argument gives
	\begin{equation}
	[c_m\left(\vec{v}_m-\vec{v}_\Gamma\right)\cdot\vec{n} ]^+_-=0.
	\end{equation}
	Conservation of momentum leads to the continuity of the normal component of the stress tensor:
	\begin{equation}
	[\tens{T} \cdot \vec{n}]^+_-=\vec{0}.
	\end{equation}
	Assuming that there are no surface dipoles or charges on the gel-bath interface, we also have continuity of the electrical potential and the displacement field;
	both follow from pillbox arguments applied to Maxwell's laws:
	\begin{gather}
		[\Phi]^+_-=0,\\
		[-\epsilon \nabla \Phi \cdot \vec{n}]^+_-=0.
	\end{gather}
	We also impose continuity of the chemical potential:
	\begin{equation}
	\left[\mu_m\right]^+_-=0.
	\end{equation}
	Furthermore, we impose the condition
	\begin{equation}\label{nopref}
	\left.\nabla c_s\right|_{\Gamma^-} \cdot \vec{n}=0,
	\end{equation}
\end{subequations}
which is only necessary on the gel side due to the derivative of the solvent concentration in the chemical potential (\ref{order}). Physically, the boundary
	condition \eqref{nopref} implies that there is no preference for solvent molecules
	to accumulate or disperse at the free surface, either of which would result
	in a local composition gradient.

Finally, we must specify a slip-type boundary condition for which there exist multiple choices that are consistent with our model. For instance,
Mori \etal\cite{Mori2013} opted for a Navier slip condition on the solvent velocity in their kinetic model of a polyelectrolyte gel. In \cite{Hennessy_debye_2020}, where we derive the electroneutral formulation of the model, we instead impose continuity of the
tangential components of the mixture velocity; however we find that the slip condition does not have a significant impact on the asymptotic analysis.

The governing equations for the gel and ionic bath, together with the coupling conditions given here at the interface, are then complemented with appropriate conditions at domain boundaries. 
In the following section we apply the framework to a specific example: a one-dimensional Cartesian geometry.
\section{Pattern formation and dynamic fluctuations}\label{application}

\begin{figure}[t]
	\centering
	\includegraphics[width=0.75\textwidth]{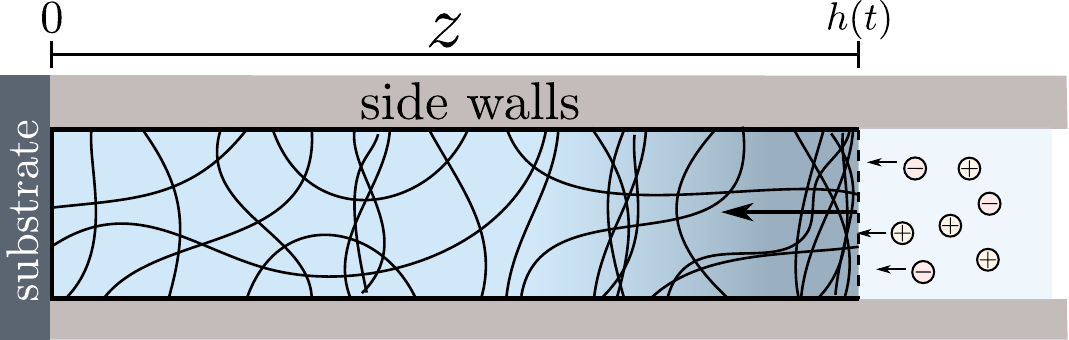}
	\caption{Sketch of a gel swelling in a ionic bath with constraints.}
	\label{1Dswell}
\end{figure}

We consider the constrained
collapse or swelling of a polyelectrolyte gel with three mobile species, namely the solvent and two ionic species, $c_+$ and
$c_-$, with opposite charges, $z_+$ and $z_-$, respectively. We model the experimental scenario in which the concentration of ionic species in the ionic bath is controlled by adding salt or pure solvent. The molecular volumes of the mobile
species are taken to be equal so that $\nu_s = \nu_{\pm} \equiv \nu$ \cite{Yu2017}. We further assume the gel is bonded to a
substrate at $z=0$ and has a free surface located at $z=h(t)$ that is in contact with an ionic bath, as shown in Figure~\ref{1Dswell}. 

We consider one-dimensional scenario in which the polyelectrolyte gel undergoes uni-axial deformations due to the uptake or release of solvent.  All velocities and fluxes have components in the $z$ direction only, and the dependent variables are functions of $z$ and time $t$ only. This reduction corresponds to the experimental scenarios~\cite{Doi2009}, where  the gel and ionic bath are free to slide along impermeable, frictionless side walls, corresponding to zero tangential stress. 
In addition, the side walls are sufficiently far apart, so that their influence on the bulk of the polyelectrolyte-ionic bath  system is negligible.

To simplify our analysis, we consider the electroneutral limit, which is justified by the smallness of the Debye length $L_d$, i.e. the length scale at which the electric field can induce charge
separation, relative to size of the gel and the bath (see Table \ref{tab:physparms}). We first discuss the homogeneous equilibrium
states obtained when all temporal and spatial derivatives, as well as all fluxes and velocities, are set to zero. We highlight that these equilibrium states are naturally electroneutral. We discuss the dependence of the bifurcation structure of the system on the material and experimental control parameters. We then consider the one-dimensional time-dependent system. We consider the system to be in a pre-swollen state, or `initial state' (see Fig.~\ref{gel_configuration}), and examine the system dynamics as we change the concentration of ionic species in the  bath, illustrating how the system transitions between equilibria. Our one-dimensional dynamic simulations predict the formation of a moving depletion front during the process of the volume phase
transition. Furthermore, by varying the salt concentration in the ionic bath, localised phase-separated structures emerge within the polyelectrolyte gel and evolve in time and space. 

\subsection{Reduction to a one-dimensional model for a constrained electroneutral gel}

\begin{table}
	\begin{center}	
		\caption{Physical parameters used in the full model 
			\eqref{full_mod_gel} and characteristic length scales of the problem.
		}
		\label{tab:physparms}
		\footnotesize
		\begin{tabular}{rll}
			\toprule[2pt]
			&\textbf{Meaning} & \textbf{Typical value(s)}  \\ \bottomrule[0.5pt]
			$k_B$ & Boltzmann's constant & $1.38\times 10^{-23}$ JK$^{-1}$  
			\\
			$T$ & Temperature & $298$ K   \\
			$e$ & Elementary charge & $1.602\times 10^{-19}$ C 
			\\
			$\epsilon$ & Absolute permittivity of bath and gel (based on water) & $7\times 10^{-10}$ Fm$^{-1}$ \\
			$\nu$  &Volume per molecule of mobile species &  $10^{-28}$ m$^3$  (\cite{Yu2017})\\
			$\mathcal{D}^{0}_i$ & Diffusivity of mobile ions in pure solvent &
			$10^{-9}$ m$^2$ s$^{-1}$ (\cite{SherwoodThomasK1975Mt})
			\\
			$\mathcal{D}_i$ &Diffusivity of mobile ions in gel& 	$\D_i=\D^{0}_i$\\
			$k$ & Hydraulic permeability of solvent in the network &$(\D_s /k_BT)\phi_n^{-\theta}$\\
			$\theta$ & Exponent in permeability law & $0, 1.5$ (\cite{Hennessy2020}) \\
			$\D_s$ &Diffusivity of the solvent in the  gel&  $\D_s=0.1\D_i$ (\cite{Drozdov2016b})
			\\
			$\chi$ &Flory interaction parameter&$0.1$--$2.5$ \\
			$C_f$ & Concentration of fixed charges in the dry gel& $\nu C_f\sim 0.01$--$0.5$\\
			$G$ &shear modulus& $10^4$--$10^5$ Pa\\
			$L$ & Typical length of a gel & 0.001--0.01 m\\
			$L_d$&  Width of the electric double layer (Debye length)& $\displaystyle L_d= \sqrt{\epsilon k_B T\nu/e^2} \sim 10^{-10}$ m\\
			$L_i$&  Width of diffuse interfaces & $\displaystyle L_i= \sqrt{\gamma/(k_BT\nu)}$\\
			$\gamma$ & Interface stiffness parameter &  Chosen so that $L_d\ll L_i<L$\\\bottomrule[2pt]		
		\end{tabular}
	\end{center}
\end{table}

In the one-dimensional scenario, the velocity and flux vectors take the form $\vec{v}=v\vec{e}_z$, where $\vec{e}_z$ is a unit vector in the direction of swelling.  Moreover,  the deformation tensor $\tens{F}$ and the stress tensor $\tens{T}$ in the polyelectrolyte gel are, respectively:
\begin{equation}\label{Fmc}
\tens{F}=\mbox{diag}\left(\lambda_0,\lambda_0,\lambda(z,t)\right), \quad \tens{T}=\mbox{diag}\left(T_\ell(z,t),T_\ell(z,t),T(z,t)\right)
\end{equation}
where $\lambda_0$ is a fixed stretch of the network in the $x$ and $y$
directions and $\lambda(z,t)$ is the axial stretch, while $T_\ell(z,t)$ is the lateral stress in the $x$ and $y$ directions while $T(z,t)$ is the stress in the axial direction.

As discussed in the companion paper \cite{Hennessy_debye_2020}, the electroneutral limit allows us to neglect the left hand-sides of Equation~\eqref{Poissongel} and~\eqref{poissonbath}. In the one-dimensional setting
considered here, the resulting equalities
\begin{subequations}   \label{en}
	\begin{alignat}{2}
		z_-c_- &= z_fc_f +z_+c_+, &\quad &0< z < h(t),\\
		z_-c_-&=z_+c_+, &\quad &z > h(t),
	\end{alignat}
\end{subequations}
for the gel and bath models, respectively, can be used to eliminate one of the ionic concentrations in each of the domains, \textit{e.g.} $c_-$ can be uniquely defined in terms of the co-ions $c_+$. In addition, the Maxwell contribution to the stress tensor $\tens{T}_M$ in Equations~(\ref{eq:Stressgel}) can be neglected in the gel. 

The electroneutrality relationships \eqref{en} hold everywhere in the bulk of the gel and bath, but break down near the interface between the two where an electric double layer develops. Resolving this `inner layer' via a perturbation expansion, starting from the interfacial conditions outlined in \S \ref{IntCond}, we obtain the following conditions at the free interface $z=h(t)$ (see \cite{Hennessy_debye_2020} for derivation details):
\begin{subequations}\label{bbc}
	\begin{align}
		\partial_z c_s(h(t)^{-},t) &=0,   \\
		T(h(t)^-,t) &=0,\\
		\mu_m(h(t)^-, t)&=\mu_m(h(t)^+, t), \quad m\in\M \label{mucont}.
	\end{align}%
\end{subequations}

We assume that the gel at $z=0$ is attached at the substrate, which is impermeable and electrically insulated. Therefore we impose
\begin{subequations}
	\begin{align}
		{v}_n(h(t)^{-},t)={0},\quad 
		{j}_m(h(t)^{-},t)=0,\quad m\in\M.
	\end{align}
	We note that this is a simplification of the possible electrical interactions between the gel and the substrate interface, where an electric double layer could also form in the scenario of the substrate not being a perfect insulator.
	To close the system we also impose the ``no-preference'' condition
	\eqref{nopref}:
	\begin{align}
		\displaystyle{\frac{\partial c_s}{\partial z}} =0.
	\end{align}
\end{subequations}

Finally, we specify far-field conditions for the  bath as follows
\begin{subequations}
	\begin{align}
		c_{+}^{(b)} \rightarrow c_0\quad\text{s.t.}\quad z_{+} c_0 + z_{-} c_{-}^{(b)} = 0,
		\label{cffc} \\
		\Phi \rightarrow 0, \quad p\rightarrow 0.
	\end{align}\label{far_cond_bath}%
\end{subequations}
Here~(\ref{cffc}) represents the scenario in which the cation concentration is controlled in the far field so as to satisfy electroneutrality, while~(\ref{far_cond_bath}) set the reference values of the electric potential and the pressure, which are taken to be zero.

\subsection{Homogeneous equilibrium solutions}\label{equil1D}

We now investigate the homogeneous steady states of the system and set all the temporal and spatial derivatives, as well as all the fluxes and velocities in (\ref{fullfinal}) and (\ref{bath_model}), to zero.  We
introduce superscripts ${(g)}$ and ${(b)}$ to distinguish variables in
the gel and  bath, respectively, denoting the homogeneous solutions by $(c_s^{(g,b)},c_+^{(g,b)},c_-^{(g,b)},\Phi^{(g,b)},p^{(g,b)}, T^{(g,b)})$. As determining
the governing equations for the homogeneous equilibria is straightforward,
we only present the main results here but provide full details in \ref{AppE}.

The homogeneous steady states satisfy the electro-neutrality condition (see equations (\ref{Poissongel}) and (\ref{poissonbath})). 
In the bath we find $p^{(b)}\equiv 0$,  $T^{(b)}\equiv0$, $\Phi^{(b)}\equiv 0$, $c_+^{(b)}\equiv c_0$ and $c_-^{(b)}\equiv z_+c_0/z_-$.
Imposing continuity of chemical potentials and normal stress at the interface, we obtain the following system of non-linear equations in terms of the three unknowns  $c_s^{(g)}$, $c_+^{(g)}$ and $\lambda$, which depend on the parameters $\mathcal{G} = G\nu/(k_BT)$, $\alpha_f=z_f\nu C_f/z_+$, $\zeta=z_-/z_+$, $\nu c_0$, $\lambda_0$ and $\chi$ as follows:
\begin{subequations}\label{group}
	\begin{align}\label{fin_eq_text}
		\mathcal{G}\frac{\left(\lambda^2-1\right)}{\lambda_0^2\lambda}+ \ln \left(\frac{\nu c^{(g)}_s}{1-\left(1-\zeta^{-1}\right)\nu c_0 }\right) +\frac{\chi(1-c^{(g)}_s\nu )+1}{\lambda_0^2\lambda}=0, \\[2.5mm]
		\left(c_+^{(g)}\right)^{1-\zeta}-\kappa^{1-\zeta}+\frac{\alpha_f}{\lambda_0^2\lambda}\left(c_+^{(g)}\right)^{-\zeta}=0,\label{fin_eq_2} \\
		\lambda_0^2\lambda-\left(1-\nu \left(c^{(g)}_s+c^{(g)}_+-\frac{\kappa^{1-\zeta}}{\zeta}\left(c^{(g)}_+\right)^{\zeta}\right)\right)^{-1}=0,
	\end{align}
	where 
	\begin{align}
		\kappa=\kappa(c^{(g)}_s,\lambda;\nu c_0,\lambda_0,\zeta)=\exp\left(\frac{\chi}{\lambda_0^2\lambda}\right) \left[\frac{\nu c_0c_s^{(g)}}{1-\left(1-\zeta^{-1}\right)\nu c_0}\right].\label{kappa}
	\end{align}	
\end{subequations}
The parameter $\alpha_f$ measures the number of
fixed charges per molecule (relative to the  valences of the
fixed to mobile species) and 
$\mathcal{G}$ is a dimensionless shear modulus.
Having determined $c_s^{(g)}$, $c_+^{(g)}$ and $\lambda$, the remaining quantities $p^{(g)}$, $T^{(g)}$,$\Phi^{(g)}$ and $c_-^{(g)}$ can be computed.
Equation~(\ref{fin_eq_2}) is derived from a generalisation of the standard Donnan equilibrium \cite{donnan_theory_1924,huyghe_quadriphasic_1997} to our specific problem (see Eq.~(\ref{DonA})), where we have an additional exponential contribution in~(\ref{kappa}) due to the mixing energy.

To further reduce the parameter space, we assume $z_+=-z_-$, so that $\zeta=-1$. Equation~\eqref{fin_eq_2} can then be solved explicitly,
\begin{equation}
c^{(g)}_+ = - \frac{\alpha_f}{2\lambda^2_0\lambda } + \sqrt{\left( \frac{\alpha_f}{2\lambda^2_0\lambda } \right)^2 + \frac{c^2_0(c^{(g)}_s)^2 \nu^2}{{\left(1-2c_0\nu\right)^2}} \exp\left(2\frac{\chi}{\lambda^2_0\lambda}\right)}, \label{cpeq}
\end{equation}
and we can rewrite \eqref{group} in terms of the two unknowns $c_s^{(g)}$ and $\lambda$:
\begin{subequations}
	\label{constctext}
	\begin{align}
		0=\mathcal{G}\frac{\lambda^2-1}{\lambda^2_0\lambda} + \ln \left(\frac{\nu c^{(g)}_s}{1-2\nu c_0}\right)  +
		\frac{\chi(1-c^{(g)}_s\nu)+1}{\lambda\lambda_0^2},\label{constc1text}\\
		\frac{\lambda\lambda^2_0-1}{\lambda\lambda^2_0}=\nu c^{(g)}_s+2\sqrt{\left( \frac{\alpha_f}{2\lambda\lambda^2_0} \right)^2 + \left(\frac{\nu c_0\nu c_s^{(g)}}{1-2c_0\nu}\right)^2 \exp\left(\frac{2\chi}{\lambda^2_0\lambda}\right)}.
	\end{align}
\end{subequations}
Equations (\ref{constctext}) reveal that the behaviour of the system depends on five (dimensionless) parameters 
$\alpha_f$, $\G$, $\lambda_0$, $\chi$ and $\nu c_0$. The first two are material
parameters. The third reflects a fixed pre-stretch of the network in the $x$ and $y$ directions (see equation (\ref{Fmc})) and is determined by the experimental setup. While for a given experiment, $\alpha_f$, $\G$, $\lambda_0$ are fixed, the remaining two parameters can be controlled during the experiment by altering
environmental conditions as follows. The Flory-Huggins parameter $\chi$ can be manipulated
by increasing or lowering the temperature or by adding certain compounds
to the  bath, such as acetone \cite{Ohmine1982,Yu2017}, while the
concentration of ions in the bath $\nu c_0$ can be set to the desired level by
adding salt or pure solvent.

\subsubsection{Results} 
\label{sec:hom_steady_state}
\begin{figure}[t]
	\centering
	\includegraphics[width=0.85\textwidth]{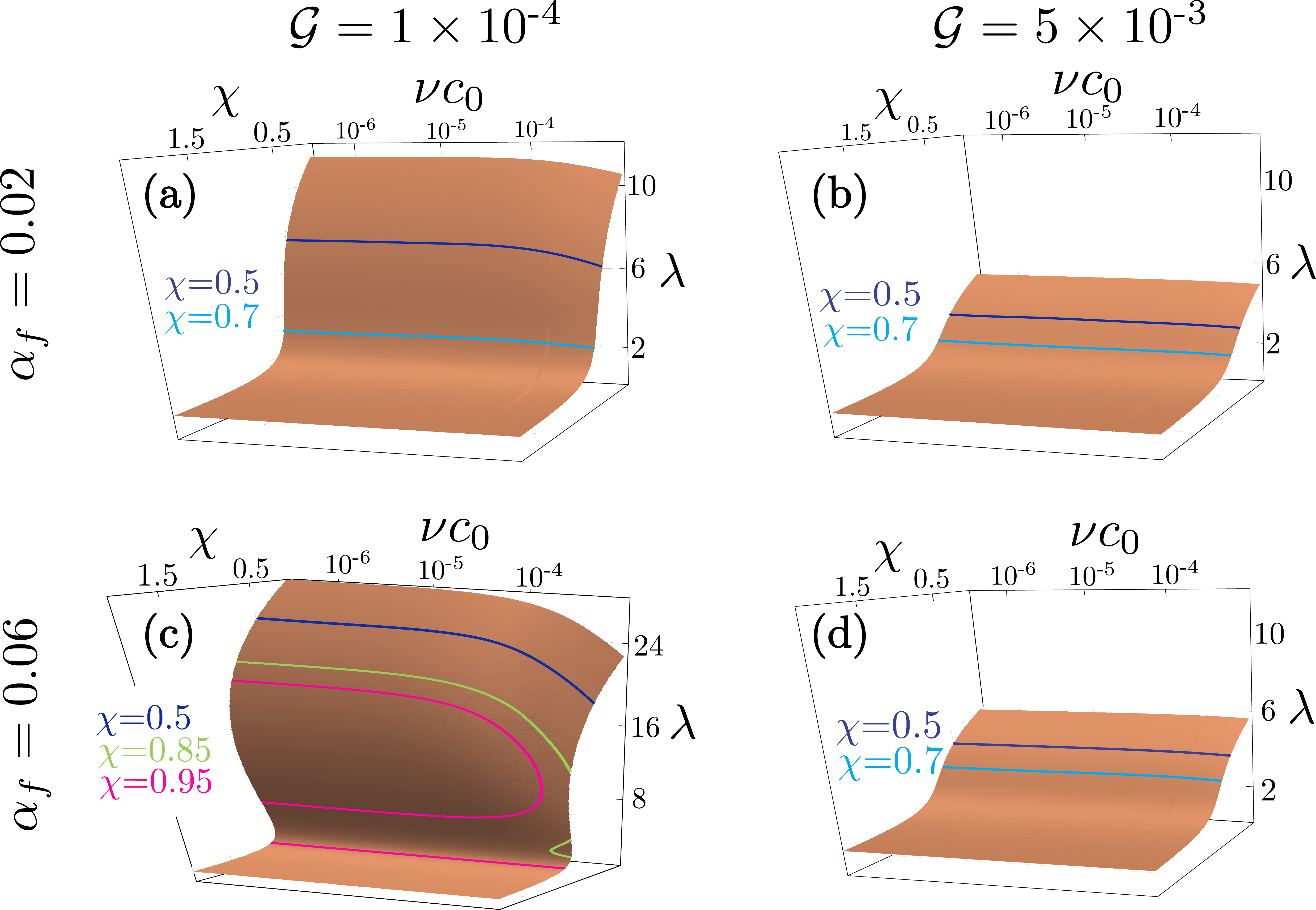}
	\caption{Equilibrium manifolds as defined by Equations~(\ref{constctext}) for a non-pre-stretched ($\lambda_0=1$) gel with different mechanical ($\G$) and electrical ($\alpha_f$) properties. We consider the domain $0.01 \leq \chi \leq 2$ and $10^{-6} \leq \nu c_0 \leq 10^{-2.5}$.}
	\label{eq_manifold1}
\end{figure}

\begin{figure}[t]
	\centering
	\includegraphics[width=0.8\textwidth]{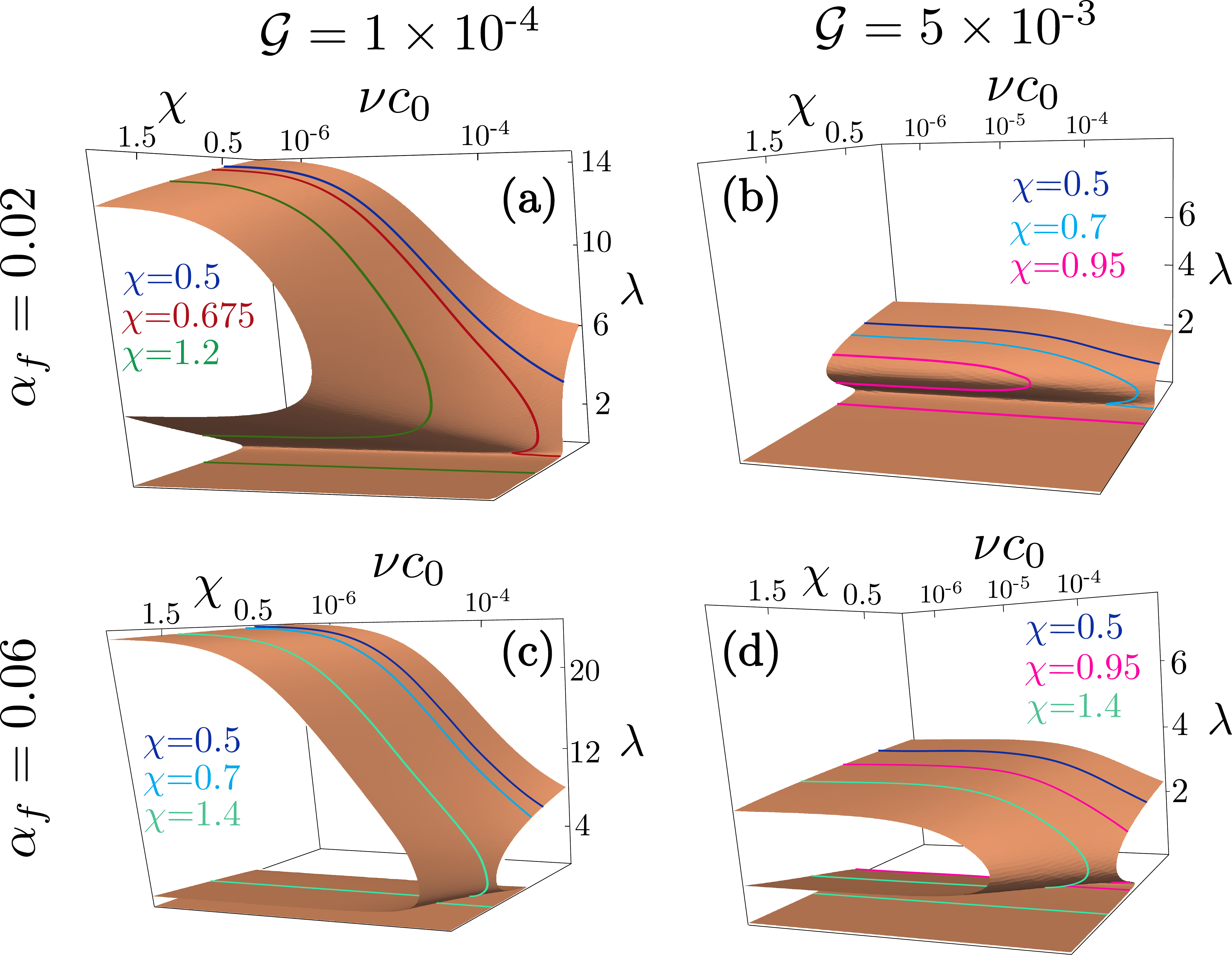}
	\caption{Equilibrium manifolds as defined by Equations~(\ref{constctext}) for a pre-stretched ($\lambda_0=5$) gel with different mechanical ($\G$) and electrical ($\alpha_f$) properties. The parameters coincide with those in Figure \ref{eq_manifold1}.}
	\label{eq_manifold2}	
\end{figure}

In Figure~\ref{eq_manifold1} and \ref{eq_manifold2},
we investigate how the equilibrium stretch $\lambda$ depends on the
ion concentration in the bath $\nu c_0$ and the Flory interaction parameter
$\chi$ for different gels and degrees of pre-stretch $\lambda_0$, as
characterised by triples $(\G, \alpha_f, \lambda_0)$. 
This amounts to computing the equilibrium manifold defined in the
space $\left(\nu c_0, \chi, \lambda\right)$  for different
choices of $(\G, \alpha_f, \lambda_0)$.  Figures~\ref{eq_manifold1} and \ref{eq_manifold2} consider the cases $\lambda_0 = 1$ (no pre-stretch) and
$\lambda_0 = 5$, respectively. 

In general, we notice that for small $\chi$ ($\chi\sim 0.5$) the gel presents a single, swollen
equilibrium i.e.\ with a large value of $\lambda$ that remains constant as the
salt concentration is increased, see e.g.\ the blue line for $\chi=0.5$ in 
Figure \ref{eq_manifold1}a.  The degree of swelling $\lambda$
decreases for larger $\G$, corresponding to stiffer gels (\emph{e.g.} compare Figures \ref{eq_manifold1}c and \ref{eq_manifold1}d). In contrast, for sufficiently large
$\chi$, the gel typically stays in the dry 
state corresponding to small values
of $\lambda$ for all salt concentrations.
In Figure~\ref{eq_manifold1}c, this behaviour is observed for values of 
$\chi$  larger than 0.95. For stiffer gels ($\G = 5\times 10^{-3}$) and/or small
concentrations of fixed charges ($\alpha_f=0.02$), see Figures~\ref{eq_manifold1}a,b,d, values of $\chi$ between these
extremes show qualitatively  the same behaviour with a flat line that shifts
to smaller $\lambda$ (less swollen gels) as $\chi$ is increased.

For softer gels with high fixed charge concentrations, the degree of swelling for intermediate $\chi$ values decreases rapidly with increasing
salt concentration; see \emph{e.g.} the $\chi=0.85$ contour in Figure \ref{eq_manifold1}c.   Moreover, for a range of lower salt concentrations
multiple solutions are possible. For example, for $\chi=0.95$, 
three solutions for $\lambda$ exist over almost the full range of $\nu c_0$ considered in the Figure \ref{eq_manifold1}c. If
we start with small $\nu c_0$ on the swollen branch, i.e.\ the branch with the
largest value of $\lambda$, and increase the salt concentration, then
the degree of 
swelling will first decrease slowly and then more rapidly. At a critical bifurcation value of $\nu c_0$, this upper branch joins with the middle branch, and for values of $\nu c_0$ beyond this point, only the lowest branch, corresponding to the collapsed state, exists.
Once on the collapsed branch, the solution will remain on it, even if
the salt concentration is decreased below the bifurcation value, and we observe the typical hysteretic
behaviour. For a given value of $\alpha_f$, we typically observe that there is
a critical value of the dimensionless shear modulus, denoted by $\G_c$,
below which multiple solutions for the stretch exist,
as seen in Figure \ref{eq_manifold1}c,d.

The behaviour of the equilibria shown in Figure \ref{eq_manifold1}
has also been observed
for freely swollen (unconstrained) polyelectrolyte gels \cite{Yu2017}. In the constrained case considered here, we have the degree of pre-stretch, $\lambda_0$, as an additional parameter. As the value
of $\lambda_0$ is increased (compare Figure \ref{eq_manifold1} and Figure \ref{eq_manifold2}), the range of $\cal G$ and $\alpha_f$ where multiple solution branches are possible increases. For a given $\alpha_f$, this corresponds to an increase in the critical value of $\mathcal{G}_c$, below which multiple solutions are obtained. Thus multiple solution branches and associated  hysteretic behaviour are possible for stiffer gels 
and for lower
concentrations of the fixed charges. 

By comparing the constrained
and the free-swelling cases directly (see \ref{app:freeswlling} for a detailed discussion of the latter), it is possible to show that while in the absence of pre-stretch ($\lambda_0=1$) the appearance of multiple solution branches in the constrained case requires softer gels than in the free-swelling case, this behaviour can be reversed by considering larger values of $\lambda_0$. 
To see this, we compare \eqref{constctext} with the
corresponding result for the free-swelling case:
\begin{subequations}
	\begin{align}
		0=
		\mathcal{G^*}\frac{{\lambda^*}^2-1}{{\lambda^*}^3}+ \ln\left(\frac{c^{(g)}_s \nu}{1-2c_0\nu}\right) +
		\frac{\chi(1-c^{(g)}_s\nu)+1}{{\lambda^*}^3}\label{free-cond1}\\
		\frac{{{\lambda^*}}^3-1}{{\lambda^*}^3}
		=\nu{c}^{(g)}_s+2\sqrt{\left( \frac{\alpha_f}{2{\lambda^*}^3} \right)^2 
			+\left( \frac{\nu c_0\nu c^{(g)}_s}{{1-2c_0\nu}}\right)^2 \exp\left(2\frac{\chi}{{\lambda^*}^3}\right)}\label{free-cond2},
	\end{align}
\end{subequations}
where $\mathcal{G}^*$ and $\lambda^*$ correspond to the stiffness and stretch of the swollen gel. 
If we assume $\lambda, \lambda^*\gg 1$ and drop the $-1$ in the numerator of the first
term in \eqref{constc1text} and \eqref{free-cond1}, the solution can be mapped from
the constrained to the free case 
via
\begin{equation}
\lambda_0^2\lambda=\lambda_*^3, \qquad {\cal G}\lambda^2={\cal G}_*\lambda_*^2,
\end{equation}
which implies that 
\begin{equation}
\frac{{\cal G}_*}{\cal G}=\frac{{\lambda^*}^4}{{\lambda_0}^4}.
\end{equation}

Suppose now we observe multiple solutions for the stretch in the freely swelling gel for values of $\mathcal{G}^*$ less than a critical value $\mathcal{G}^*_c$.  
Then, for $\lambda_0=1$, we observe that the corresponding critical value of stiffness for the constrained gel is such that $\mathcal{G}_c < \mathcal{G}^*_c$, and multiple solution branches for the stretch appear at softer gel stiffnesses for the constrained case compared with the freely swelling case. If, however, the pre-stretch is much larger so that $\mathcal{G}_c > \mathcal{G}^*_c$, we obtain the opposite behaviour with multiple solution branches observed in the constrained case for stiffer gels compared with the freely swelling case. This agrees with observations in the literature
\cite{Horkay2001,duskova-smrckova_how_2019} that volume phase
transitions in compressed gels occur earlier, e.g.\ at lower values of the salt
concentration, and vice-versa for decompressed (stretched) gels. 

\subsection{Swelling and collapse dynamics}
\label{sec:dynamicresults}

Many theoretical studies of the volume phase transition in
polyelectrolyte gels focus on equilibrium states. 
Similarly, experiments often report on the final states as well as those
aspects of the dynamics that are slowly evolving.
Our model allows us to investigate the entire range of transient dynamics,
which include fast waves of mobile ions and the slow evolution of
interfaces between phase-separated regions that appear during gel collapse.

With this in mind, we now present time-dependent simulations of the simplified one-dimensional model, and illustrate how the system transitions between the equilibrium states identified in \S\ref{equil1D}. In the analysis that follows, we assume the gel is not pre-stretched (i.e. $\lambda_0=1$) and that the ionic salt to be monovalent (i.e. $z_\pm=\pm1$). Further assuming the bath is very large (so that small changes in ion concentration due to ion exchange with the gel can be neglected), we see that the bath maintains its equilibrium state,  characterised by constant ionic concentrations, $c_\pm^{(b)}=c_0$. 

We non-dimensionalise the governing equations as follows
\begin{equation}
\begin{aligned}
\mu^*_m = \frac{\mu_m-\mu^0_m}{k_BT}, \quad \phi_m = \nu c_m, \quad \phi_f=\nu c_f, \quad \Phi^* = \frac{\Phi e}{k_B T},  \\
\tens{T}^*=\frac{\tens{T}}{G}, \qquad z^* =\frac{z}{L}, \quad t^*=\frac{t}{\tau}, \quad p^*= \frac{p}{G}, \quad j^*_m=\frac{\nu L}{\D_s}j_m,
\end{aligned}
\end{equation}
so as to focus on the length scale of the gel $L$ and the time scale of the solvent diffusion $\tau=L^2/\D_s$. 
The resulting dimensionless system of governing equations for the volume fractions $\phi_m$, together with the boundary and interfacial conditions,  are presented in \S2 of Celora et al.~\cite{Celora_siap_2020}. 
In addition to the dimensionless material parameters highlighted in \S\ref{equil1D} ($\mathcal{G}$ and $\alpha_f$), we now have two additional parameters, fixed for a given experimental scenario,
\begin{equation}
\D^*_\pm=\frac{\D_\pm}{\D_s}, \quad \omega= \sqrt{\frac{\gamma}{\nu k_BTL^2}},
\end{equation}
where $\D^*_\pm$ are the relative diffusivities of the ions with respect to the solvent and $\omega$ is given by the ratio of the interfacial and the gel length scales (see respectively $L_{i}$ and $L$ in Table \ref{tab:physparms}). We expect $\omega$ to be small, therefore, given the lack of precise estimate, we fix its value to  $\omega=2.5\times 10^{-2}$. Following \cite{Hennessy2020}, we
set the exponent in the permeability law in \eqref{k_def}
to be $\theta = 0$. 
In line with reported values from the literature (see Table \ref{tab:physparms}) we set $\D^*_\pm=10$. 

\subsubsection{Results}

We first consider the case of a volume phase transition discussed by Yu et al. \cite{Yu2017},
who observe good agreement between their free-swelling equilibrium solutions and the experiments
by Ohmine and Tanaka \cite{Ohmine1982}. In particular, their theory captures the parameter
values for which a phase transition occurs. These are, in our notation, $0.98<\chi<1.55$, $\G=1.09\times10^{-3}$, $0.02<\alpha_f<0.1$, $0<\nu c_0<0.06$. 
The parameter values in Figure~\ref{eq_manifold1} lie in this range, with a tendency towards
softer gels or smaller values of $\chi$ to compensate for the fact that in the constrained
swelling scenarios, volume phase transitions are shifted. 

We now present some examples of the complex dynamics that can be observed by changing the concentration of ions in the bath. These perturbations drive the evolution of the gel the system from
one equilibrium state to another. As we will show, such dynamics can be accompanied by phase separation that nucleates at the free surface and forms an interface between solvent-rich (swollen) and solvent-poor (collapsed) phases that then propagates through the gel.
 
\begin{figure}
	\includegraphics[width=\textwidth]{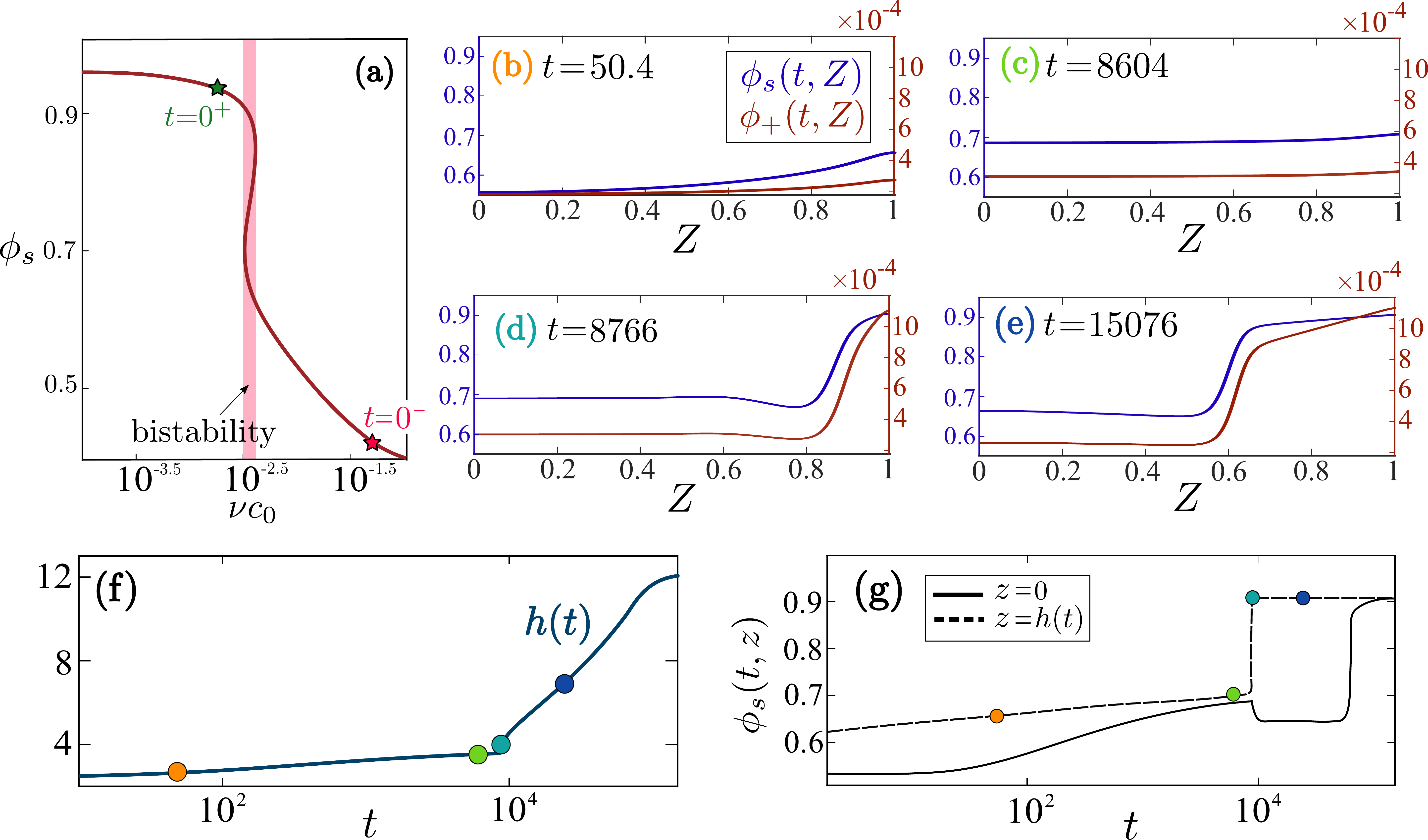}
	\caption{Swelling of a soft ($\G=10^{-4}$) gel with a high concentration of fixed charges ($\alpha_f=0.1$) in contact with an ionic bath. The gel is initially in equilibrium with the bath and $\nu c_0=0.05$. At time $t=0$ the concentration of ions in the bath is decreased to $\nu c_0=10^{-2.75}$. (a) Equilibrium curve for the gel for $\chi=0.95$. We highlight the initial and the final solvent fraction in the gel (stars) and the bistability region (shaded in red). (b-f) Snapshots of the volume fractions of solvent and co-ions in the gel at different times and the evolution of the gel size $h(t)$. The spatial variable $Z$ is defined as $Z=z/h(t)$.  (g) Evolution of the volume fraction of solvent at the two boundaries of the gel, i.e. $z=0$ and $z=h(t)$.}
	\label{simulation1}
\end{figure}

In the first case, we simulate the swelling of a gel that is triggered by
decreasing the concentration of ions in the bath $\nu c_0$. In Figure \ref{simulation1}a we show the homogeneous equilibrium solvent volume fraction in the gel for fixed values of $\chi, \mathcal{G}$ and $\alpha_f$. There is an interval of values for $\nu c_0$ for which system is bi-stable. We choose our initial state (denoted by a red star) to be on the right of this region by setting $\nu c_0 =0.05$. Then, at time $t=0^+$, the concentration in the bath is suddenly decreased to $\nu c_0 = 0.0018$ so as to drive the gel to the swollen state just to the left of the bistable region (denoted by a green star).

The dynamics of the transition between the two equilibrium states is shown in Figure \ref{simulation1}(b-f), which shows a sequence of snapshots for the distribution of solvent and mobile ions in the gel, and the evolution of the gel size $h(t)$. Note that the spatial variable is rescaled so that results are mapped onto a fixed domain $Z=z/h(t)$. We identify two time scales. Initially the solvent diffuses into the gel and the concentration of both solvent and ions is fairly uniform (see panels (b)-(c) in Figure \ref{simulation1}), and the gel swells at a low rate (see panel (f)). However, at a later time ($t>10^4$), near the boundary with the bath, a highly-swollen region forms, hence driving the formation of a front which later propagates into the gel (see Fig.~\ref{simulation1}d) with the gel swelling now at a faster rate. The front separates two homogeneous (but not equilibrium) states. As analysed in more detail in \cite{Celora_siap_2020} via phase-plane analysis, the solvent (and ionic) volume fraction ahead of the front (from $z=0$ to the front location) $\phi_s^{(1)}$ ($\phi_+^{(1)}$) and behind the front (from the location of the front to $z=h(t)$) $\phi_s^{(2)}$ ($\phi_+^{(2)}$) are determined by a \textit{Maxwell condition} for the co-existence of the two phases. Since the volume fraction of solvent prior to phase separation is greater than $\phi_s^{(1)}$, there is an initial back-flow of solvent from the bulk to the free boundary of the gel (see the continuous black curve in Figure \ref{simulation1}g at $t\approx 10^4$). When looking at the time evolution of the gel size $h(t)$, we can clearly identify the time at which phase separation occurs as it corresponds to a sharp increase in the rate at which $h(t)$ grows. Such behaviour resembles that observed for neutral gels \cite{Tomari1995, Hennessy2020}, where phase separation was induced by changing the temperature or by forcing solvent into the gel. In this case the separation occurs naturally by tuning the concentration of ions in the bath.  

This example highlights the rich range of time and spatial scales that drive the formation of organised structures, such as the propagating wavefront into the gel. In addition, it leads to intriguing questions about the physical mechanisms responsible for selecting the wave speed and the slow/fast dynamics that can be further studied using multi-scale computational and asymptotic techniques. 

\begin{figure}
	\includegraphics[width=\textwidth]{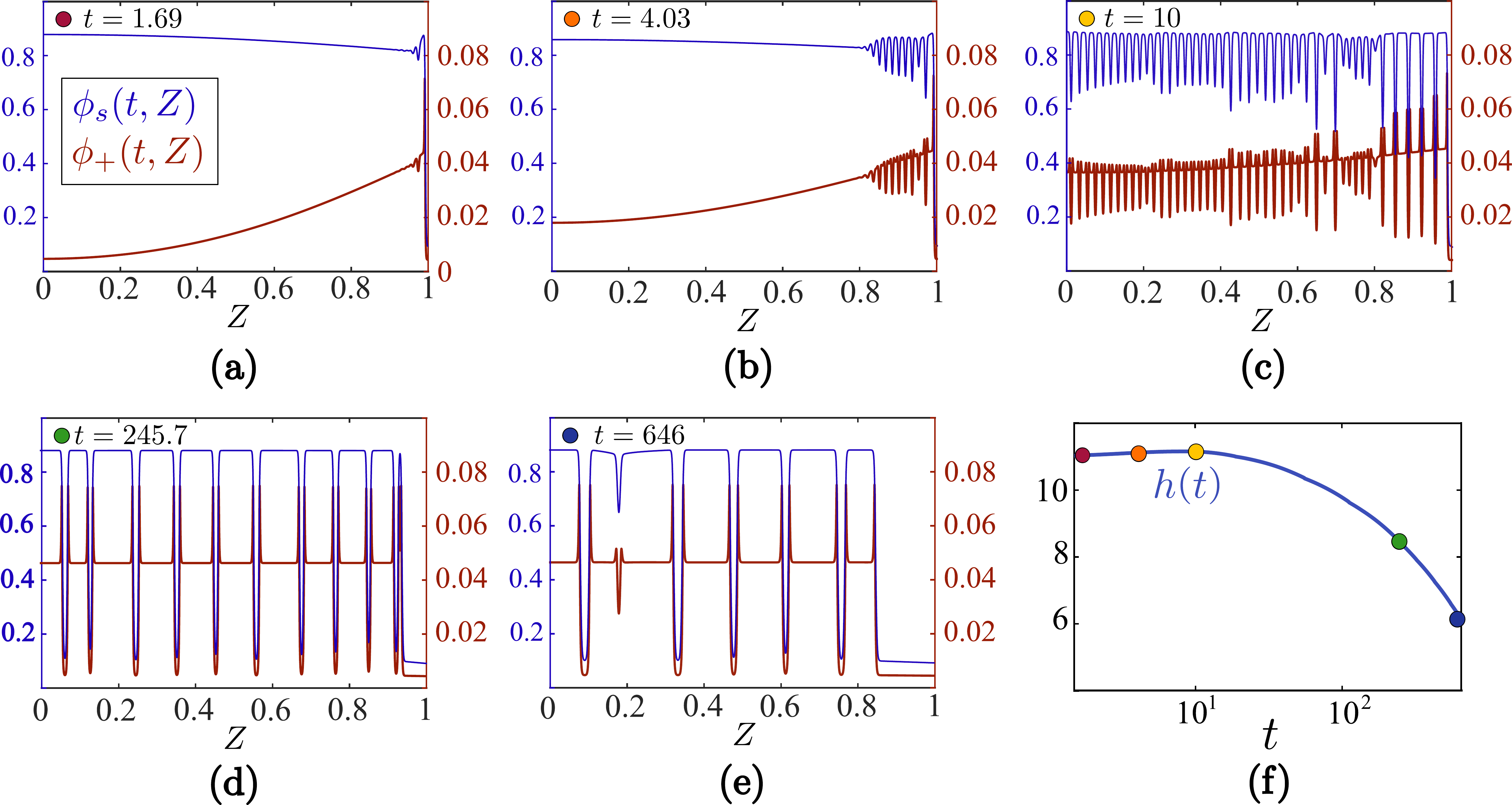}
	\caption{Spinodal decomposition in the bulk of the gel and front propagation induced by perturbing the environmental conditions. The gel is initially in equilibrium with the bath ($\nu c_0=10^{-4}$) with $\chi=1.75$. At time $t=0^{+}$ the concentration of ions in the bath is increased to $\nu c_0=5\times10^{-2}$ and $\chi$ is increased to $\chi=2.4$. The other parameters are set to: $\G=10^{-3}$, $\alpha_f=0.25$. The spatial variable $Z$ is defined as $Z=z/h(t)$.}
	\label{simulation2}
\end{figure}

Another scenario that we can investigate with our phase-field approach is microphase separation. This has been widely investigated in particular through
experiments \cite{matsuo_patterns_1992,Dusek1968}, where
the initial formation small localised regions of concentrated phases has been observed.  
While previous theoretical studies~\cite{wu_control_2010,wu_pattern_2012} concern the equilibrium states and their stability, we can now use our approach to investigate the formation and coarsening of such microphases and look at how this impacts the overall dynamics of the gel. As shown in Figure \ref{simulation2}, by changing the environmental conditions, we can induce phase separation at the free interface, as well as spinodal decomposition in the bulk of the gel. 
In these simulations, the gel is initially in equilibrium with the bath with
$\nu c_0=10^{-4}$ and $\chi = 1.75$. Then, at time $t=0^{+}$, the ionic concentration in the bath is increased to $\nu c_0=5\times 10^{-2}$ and $\chi$ is increased to $\chi=2.4$. We also take $\mathcal{G}=10^{-3}$ and $\alpha_f=0.25$. 
These parameter values are chosen to be in line with those
of Wu et al.~\cite{wu_control_2010}.

Figure~\ref{simulation2}f reveals that the gel undergoes a temporary period of swelling before ultimately
shrinking and tending towards to solvent-poor collapsed state.
This non-monotonic behaviour results from the fast diffusion of ions relative to the solvent. On short time
scales, there is a rapid influx of cations while the total solvent content
remains constant, thus leading to an increase in gel volume. However, on longer
time scales, solvent is ejected from the gel leading to its collapse. These observations  are in line with the results reported in \cite{Zhang2020}, where, for large ion concentrations, the authors also observed a fast initial phase where the size of the gel increases; after this first transient, the solvent is driven out of the gel which therefore starts to shrink in size. However, these authors attribute the non-monotonic evolution of the
gel size to Stefan--Maxwell diffusion.

The novel mechanism we are able to capture is the formation of localised collapsed phases in the highly swollen region of the gel. These coarsen, first quickly, then more slowly, as the front invades the gel until it has completely collapsed.
Moreover, we are able to observe how the electronic structure of the gel
evolves during phase separation. The results in Figure \ref{simulation2}
show that the concentration of cations (anions) is the lowest (highest) in the
collapsed phase. Interestingly, the highest concentration of cations is not
in the swollen phase, but in the diffuse interface which separates swollen and collapsed phases. As we explain in detail in \cite{Celora_siap_2020}, depending
on the parameter setting, various combinations of front propagation and
spinodal decomposition can occur en route to collapse.

\begin{figure}[htb]
	\centering
	\includegraphics[width=\textwidth]{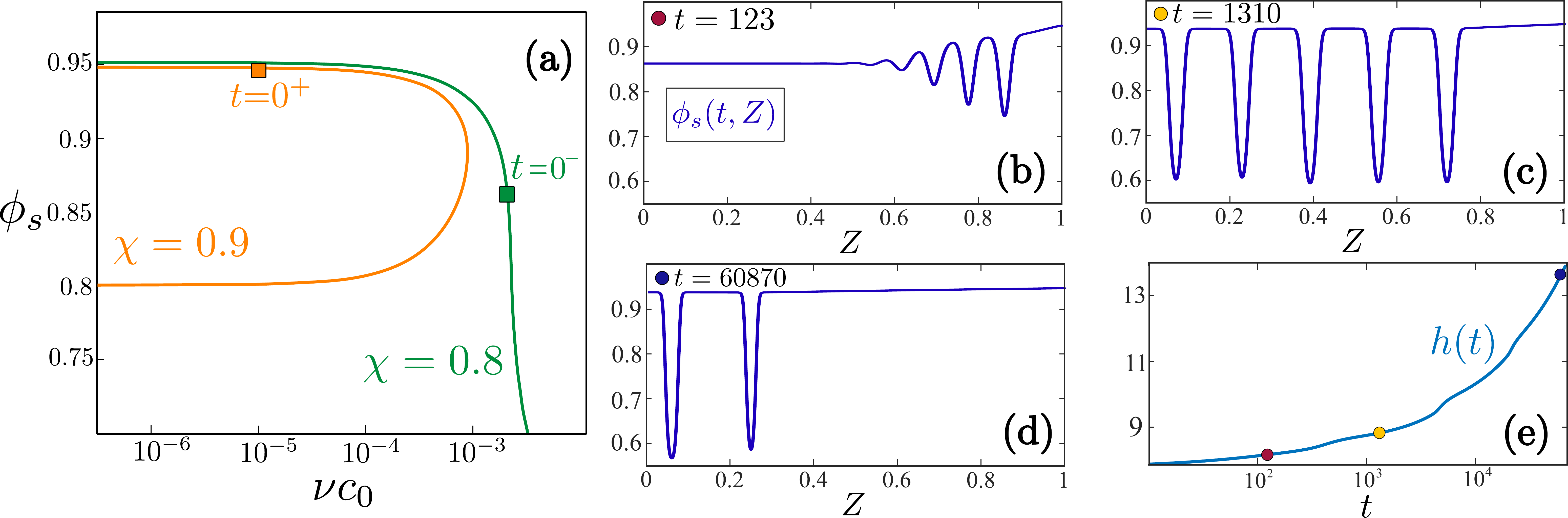}
	\caption{(a) Sections of the equilibrium manifold in Figure \ref{eq_manifold1}c for $\chi=0.8$ (green) and $\chi=0.95$ (orange line; we do not show the collapsed branch). Starting from a swollen gel (green square), we are able to drive instabilities in the bulk by increasing $\chi$ (from $0.8$ to $0.9$) and decreasing the concentration of ions in the bath (from $5\times 10^{-2}$ to $10^{-5}$). (b)--(e) The dynamics of spinodal decomposition. The simulation are interrupted prior to reaching the new equilibrium. The values of $\G = 10^{-4}$ and $\alpha_f = 0.06$ are as in Figure \ref{eq_manifold1}c. }
	\label{fig:final_sim}
\end{figure}

Spinodal decomposition can also occur in the case of a swelling experiment. By tuning the values of the parameters, as in Figure \ref{fig:final_sim}, it is  possible to drive the gel to a new highly swollen equilibrium state. In doing so, the system is pushed into the unstable (spinodal) region of the phase diagram (see \cite{Celora_siap_2020} for the full stability analysis) thus driving the formation of localised, solvent-depleted zones.

We can use our model to examine the generation of lateral stresses within
the gel, given here in dimensional units as:
\begin{equation}
T_\ell(t,z)= G\left[\frac{\phi_n^2-1}{\phi_n}+ \frac{\omega^2}{\G} (\partial_z \phi_s)^2\right].
\label{eqn:T_ell}
\end{equation}
The two terms on the right-hand
side of \eqref{eqn:T_ell} represent the contributions from the elastic and
Korteweg stresses, respectively. In the electro-neutral limit, the contributions from the Maxwell stress tensor are negligible.

\begin{figure}
	\centering
	\includegraphics[width=0.5\textwidth]{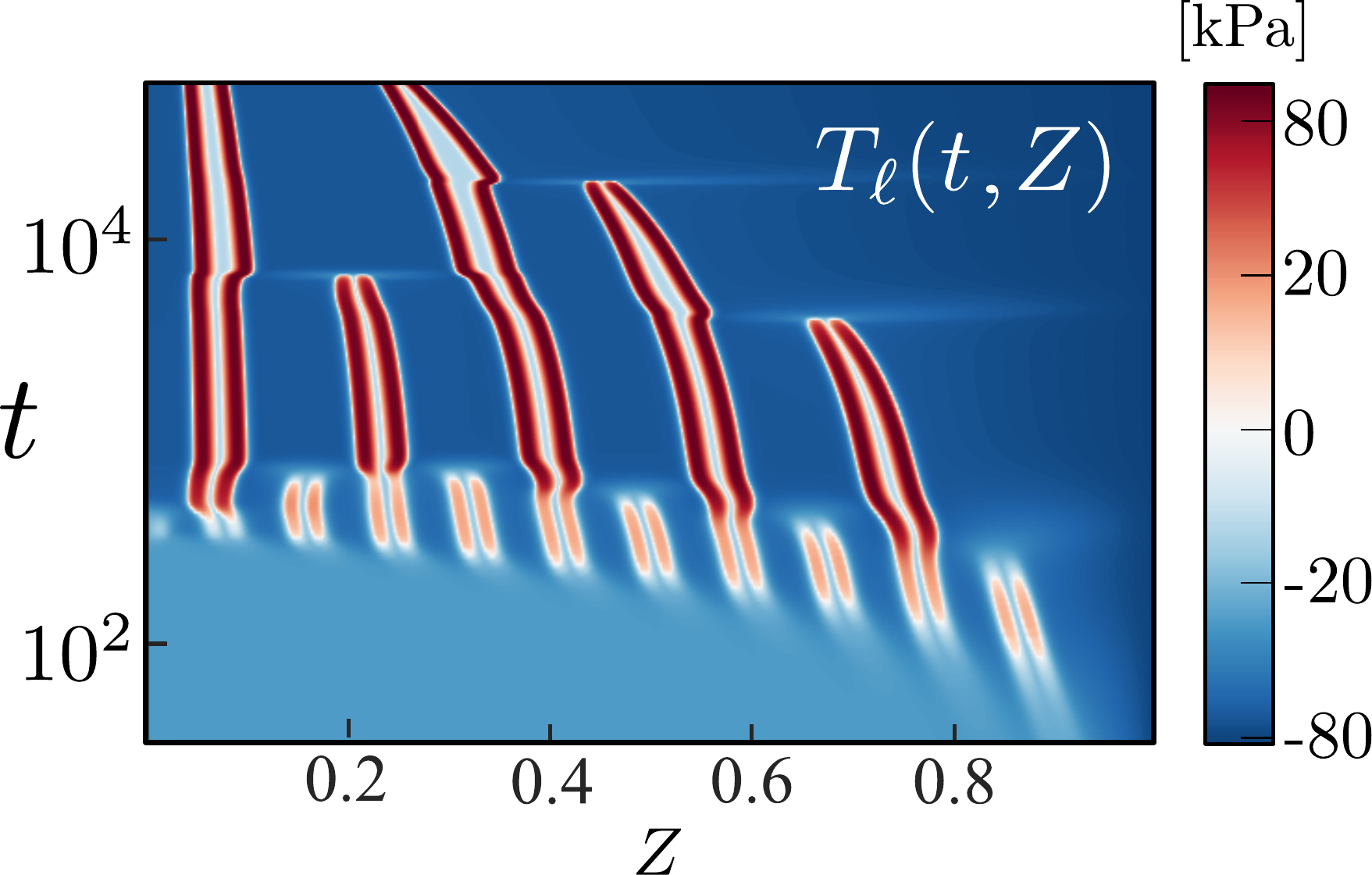}
	\caption{Time evolution and spatial distribution of the lateral stresses $T_\ell$ in the gel for simulations in Figure \ref{fig:final_sim}.}
	\label{stress_sim3}
\end{figure}

The spatiotemporal evolution of the lateral stresses are shown in Figure \ref{stress_sim3}, which have been computed using the simulation results of Figure \ref{fig:final_sim}. The bulk of the gel is initially weakly compressed compared to the highly swollen region behind the front (near $Z\approx1$). The initiation of spinodal decomposition has a two-fold impact on the stress distribution within the gel. Firstly, it rapidly increases the compressive stress experienced by the bulk of the gel,
which evolves towards a state of larger solvent content; see Figure \ref{fig:final_sim}b--d. Secondly, it locally relaxes the compression in the gel via the formation of thin and isolated solvent-depleted (or collapsed) phases. The generation of a highly non-uniform compressive stress might drive the onset of instabilities similar to those observed by Matsuo and Tanaka~\cite{matsuo_patterns_1992}. The diffuse interfaces separating the collapsed and swollen phases within the gel experience a large tensile stress due to the contribution to $T_\ell$ arising from the
Korteweg stress. The non-trivial distribution of the stresses suggests the emergence of complex patterns that can be investigated by considering a full two- or three-dimensional  geometry. 

\section{Conclusions and outlook}

We have derived a time-dependent phase-field model for a polyelectrolyte gel surrounded by an ionic bath.
The governing equations for the gel and bath are coupled through boundary
conditions at the gel-bath interface. Near this interface, we expect the
formation of an electric double layer, which is the focus of our companion
paper \cite{Hennessy_debye_2020}.
The systematic model derivations are based on linear non-equilibrium thermodynamics, which allow for multi-component transport of ions and solvent to be described in terms of Stefan--Maxwell diffusion.
In addition to the non-linear elasticity of the gel, we also account for the free energy of internal interfaces associated with phase separation to capture the transient dynamics of patterns that form during gel swelling and collapse.
The resulting comprehensive model opens the doors to new investigations into  the emerging patterns and their dynamics resulting from the interplay of the underlying physics interacting on multiple time and spatial scales. In particular, we can exploit the model to  probe and predict the impact of environmental changes or stimuli on the structural transitions of the polyelectrolyte gel.

In a first set of investigations, we formulate the boundary value problem corresponding to constraining the gel in a one-dimensional setting.  We derive the equilibrium manifolds and their bifurcations as a function of the shear modulus of the gel, the concentration of fixed charges of the polyelectrolyte and the Flory-Huggins interaction parameter, and discuss the impact of an applied stretch and changes in the salt concentration in the bath on the bifurcations of the system and the existence of multiple solution branches. Additionally, we present some dynamical simulations for one-dimensional constrained swelling, where we resolve the transient dynamics of the gel collapse/swelling. Small changes in the salt concentration set off an initial fast wave in ion concentration that triggers a depletion front which then propagates into the swollen bulk from the free surface where the collapsed phase first appears. Besides resolving the fast dynamics, our model also allows us to describe scenarios for other (slower) salt diffusion regimes and their impact on phase transitions within the gel. 

We have further shown that the macro-scale phase separation in the gel via front propagation can be accompanied with micro-phase separation in the bulk of the gel~\cite{shibayama_spatial_1998, wu_control_2010, wu_pattern_2012}.
Interestingly, our simulations revealed that the cation concentration is the smallest in the locally collapsed phases and maximum in the
diffuse interfaces that form between collapsed and swollen phases. 
While these initial dynamics may be difficult to experimentally observe, our simulations show that these regions quickly coarsen and give rise to alternating zones of large and small compressive stress. In higher dimensions, this highly non-uniform stress distribution is likely to trigger instabilities leading to complex patterns as previously observed in experiments, for example, by Tanaka \cite{Tanaka1978} and more recently by Shen et al~\cite{Shen2019} and Chang et al~\cite{Chang2018}.

A more comprehensive analysis of the one-dimensional scenario is presented in the companion paper \cite{Celora_siap_2020}, where we present a stability and phase-plane analysis, to predict, describe and understand the formation of the patterns  observed. Even though we are able to identify the signatures of the dynamics that give rise to pattern formation in a polyelectrolyte gel, our numerical and analytical results are currently in one dimension only, which limits their experimental replication, given the intrinsic three-dimensional nature of the mechanical stress experienced by the gel. Higher dimensional extensions are part of our ongoing research.

We have also shown that polyelectrolyte gels can be driven to a more-swollen state by simply decreasing the concentration of ions in the bath. If, as shown in Figure \ref{fig:final_sim}, the Flory parameter $\chi$ is simultaneously increased appropriately, then this can lead to the development of instabilities in the bulk of the swelling gel.
These instabilities may drive the system into the region of attraction of the collapsed state, hence inverting the behaviour of the gel from swelling to collapsing. This intriguing behaviour suggests the lower branch of the bifurcation to be energetically favourable. We postpone a more rigorous analysis into the intricate balance of the physical mechanisms that are driving these transitions to future work. 

Finally, we point out that the framework we have developed, that fully resolves the dynamics of polyelectrolyte gels, paves the way for new emerging fields. For example new scenarios, such as the potential impact of shear flow on the gel can now be investigated, since now we can capture the hydrodynamics of the ionic bath accounting for its mechanical interactions with the gel. 

\appendix
\section{Conservation of mass equation in current configuration}\label{conscurrent}
Here we derive the  conservation of mass equation in the current configuration, starting from the corresponding equation in the reference configuration (\ref{consmass}).  

The integral form of (\ref{consmass}) is given by
\begin{equation}
\frac{d}{dt}\int_{\mathcal{V}_R}C_m dV_R=-\int_{\mathcal{S}_R} \vec{J}_m\cdot{\vec{N}} dS_R,
\end{equation}
where $\mathcal{V}_R$ is an arbitrary control volume in the reference configuration, $\mathcal{S}_R$ is the associated surface area with outward unit normal vector $\vec{N}$. 
This integral form can be converted to the current configuration using the relations (\ref{elem}), together with the Reynolds transport theorem, to give
\begin{equation}
\int_{\mathcal{V}(t)}\left(\frac{\partial c_m}{\partial t} +\nabla \cdot (c_m\vec{v}_n)\right) dV=-\int_{\mathcal{S}(t)} \vec{j}_m\cdot{\vec{n}}\, dS,\label{consmass_current_app1}
\end{equation}
where $c_m=C_m/J$ and $\vec{j}_m=J^{-1} \tens{F}\vec{J}_m$ are the concentration and the flux in the current configuration, respectively. Here $\nabla$ denotes the gradient with respect to the current state, and  $\vec{v}_n=\partial\vec{u}/\partial t +(\vec{v}_n\cdot\nabla)\vec{u}$ is the network velocity in the current state.  
The local balance law in Eulerian coordinates is then
\begin{align}
	\frac{\partial {c}_m}{\partial t} + \nabla \cdot (c_m \vec{v}_n+\vec{j}_m)=0. \label{mass_cons3}
\end{align}

\section{Polyelectrolyte gel equations in the reference configuration}\label{eqref}

The main text gives the governing equations for the polyelectrolyte gel in the current configuration as this largely simplifies the form of the stresses. For completeness, here we present the corresponding governing equations in the reference configuration. 
\begin{subequations}\label{fullfinal_ref}
	\begin{align}
		J=1+\sum_m\nu_m C_m,\\
		\partial_t C_s + \nabla_R \cdot \vec{J}_s = 0,\\
		\partial_t C_i + \nabla_R \cdot \vec{J}_i = 0,\ \ i\in \mathbb{I},\\
		\nabla_R \cdot \tens{S}=\mathbf{0},\\
		-\epsilon\nabla_R^2 \Phi= e\left(\sum\limits_{i\in\mathbb{I}} z_i C_i+z_f C_{f}\right),
	\end{align}
	where the nominal diffusive fluxes are
	\begin{align}
		{\vec{J}}_s =-K \tens{C}^{-1}\left(C_s\nabla_R \mu_s +\sum_i \frac{\D_i}{\D^0_i} C_i \nabla_R \mu_i\right),\\
		{\vec{J}}_i = -\frac{\D_i}{k_B T}C_i\tens{C}^{-1}\nabla_R \mu_i +\frac{\D_i}{\D^0_i} \frac{C_i}{C_s} \vec{J}_s,\ \ i\in\mathbb{I}.
	\end{align}
	The micro-stresses, chemical potentials, and osmotic pressures
	are given by
	\begin{align}
		\vec{\xi}_s = 2\gamma_1 \,\tens{C}^{-1} \,\nabla_R \,C_s-\gamma_3 \,\tens{C}^{-1} \,\nabla_R \,J,\label{sys1}\\
		\vec{\xi}_J = 2\gamma_2  \,\tens{C}^{-1} \,\nabla_R \,J-\gamma_3  \,\tens{C}^{-1} \,\nabla_R \,C_s,\\
		\mu_s = (p+\Pi_s) \nu_s + \mu_s^0 +\mu_s^G 
		,\label{gov1}\\
		\Pi_s= \frac{k_BT}{\nu_s}\left[\ln \left(\frac{C_s \nu_s}{J}\right)+1-\sum_{m\in\mathbb{M}} \frac{C_m\nu_s}{J} +
		\frac{\chi(J-C_s\nu_s)}{J^2}\right],\\
		\begin{aligned}
			\mu^G_s =-\nabla_R\cdot\boldsymbol{\xi}_s+G_{iJ}G_{iM}\left( \frac{\partial \gamma_1}{\partial C_s}\frac{\partial C_s}{\partial X_J}\frac{\partial C_s}{\partial X_M}+\frac{\partial \gamma_2}{\partial C_s}\frac{\partial J}{\partial X_J}\frac{\partial J}{\partial X_M}-\frac{\partial \gamma_3}{\partial C_s}\frac{\partial C_s}{\partial X_J}\frac{\partial J}{\partial X_M}\right),\end{aligned}\\[3pt]
		\mu_i = (p+\Pi_i) v_i + \mu^0_i + e\Phi z_i,\quad i\in\I,\label{mu}\\
		\Pi_i=\frac{k_BT}{\nu_i} \left[\ln \left(\frac{\nu_iC_i}{J}\right)+1-\sum_{m\in\mathbb{M}}\frac{\nu_iC_m}{J} -\frac{\chi C_s\nu_i}{J^2} \right],\quad i\in\I.
	\end{align}
	The nominal stress tensor as well as the elastic and Maxwell contributions are
	\begin{align}
		\tens{S}= -pJ \tens{F}^{-T} +  \tens{S}_K +\tens{S}_M + \tens{S}_e,\\[2pt]
		\tens{S}_e=G\left(\tens{F}-\tens{F}^{-T}\right),\\
		\tens{S}_M=- \frac{1}{\epsilon J} \left(\frac{1}{2} \,|\tens{F}\vec{H}|^2 \tens{I} -(\tens{F}\vec{H}) \otimes (\tens{F}\vec{H})\right)\tens{F}^{-T},
	\end{align}
	Finally, the nominal Korteweg stress tensor is
	\begin{align}
		\label{sys3}
		\begin{aligned}
			\tens{S}_K\tens{F}^{T}=J\frac{\partial \gamma_1}{\partial J}G_{iJ}G_{iM}\frac{\partial C_s}{\partial X_J}\frac{\partial C_s}{\partial X_M}\tens{I}-2\gamma_1(\tens{F}^{-T}\nabla_R C_s) \otimes (\tens{F}^{-T}\nabla_R C_s)\\	
			-J\frac{\partial \gamma_3}{\partial J}G_{iJ}G_{iM}\frac{\partial C_s}{\partial X_J}\frac{\partial J}{\partial X_M}\tens{I}+2\gamma_3\, \text{Sym}[(\tens{F}^{-T}\nabla_R J) \otimes (\tens{F}^{-T}\nabla_R C_s)]\\ 
			+J\frac{\partial \gamma_2}{\partial J}G_{iJ}G_{iM}\frac{\partial J}{\partial X_J}\frac{\partial J}{\partial X_M}\tens{I}-2\gamma_2(\tens{F}^{-T}\nabla_R J) \otimes (\tens{F}^{-T}\nabla_R J) -J (\nabla_R\cdot \vec{\xi}_J)\tens{I} ,
		\end{aligned}
	\end{align}%
\end{subequations}
where $\gamma_1=\gamma/(2J)$, $\gamma_2=C_s^2\gamma/(2 J^{3})$, $\gamma_3=\gamma C_sJ^{-2}$, Sym$[\cdot]$ denotes the symmetric part of a tensor, $\tens{C}=\tens{F}^T\tens{F}$ is the right Cauchy-Green deformation tensor, and $\tens{G}=\tens{F}^{-T}$. Note that in taking the partial derivatives of $\gamma_{1,2,3}$ with respect to $C_s$ and $J$ we consider the latter two to be independent.

To move to the formulation in the current state, we use the following identities:
\begin{subequations}
	\begin{align}
		\nabla_R \cdot \vec{\xi}_s = \gamma J \nabla \cdot \left(J^{-1}\nabla c_s\right) = \gamma \nabla^2 c_s - \frac{\gamma}{J^2}\nabla C_s\nabla J + \frac{\gamma C_s}{J^3} \left|\nabla J\right|^2, \\
		\nabla_R \cdot \vec{\xi}_J = -\gamma J\nabla \cdot \left(\frac{c_s}{J}\nabla c_s\right)	=-\gamma c_s|\nabla c_s|^2- \gamma |\nabla c_s|^2
		+\frac{\gamma C_s}{J^3} \nabla C_s \nabla J-\frac{\gamma C^2_s}{J^4} |\nabla J|^2, \\
		\frac{1}{2}J\gamma \nabla c_s \otimes \nabla c_s = \gamma_1 \nabla C_s \otimes \nabla C_s-\gamma_3\text{Sym}\left[\nabla C_s \otimes \nabla J\right]+\gamma_2\nabla J \otimes \nabla J,
	\end{align}
	so that the chemical potential and the expression for tensor $\tens{S}_K$ simplify to:
	\begin{align}
		\mu_s = (p+\Pi_s) \nu_s + \mu_s^0 -\gamma \nabla^2 c_s, \\
		\begin{aligned}
			J^{-1}\tens{S}_K\tens{F}^{T}=\left(\gamma c_s|\nabla c_s|^2+\frac{\gamma}{2} |\nabla c_s|^2\right)\tens{I}- \gamma \nabla c_s \otimes \nabla c_s.
		\end{aligned}
	\end{align}
\end{subequations}

\section{Rate of mechanical and electrical work for ionic bath}\label{AppIonic}
Here we give additional details required to move between the first and second expressions for the rate of electric work on a reference volume of the ionic bath in the current state (see equation~(\ref{electricalionic})).  Starting from the first line of equation~(\ref{electricalionic}) we have 
\begin{subequations}
	\begin{align}
		\mathcal{W}_{el}(\mathcal{V}(t)) = -\int_{\mathcal{S}(t)} \Phi \left(\frac{D\vec{h}}{Dt}+\vec{h}\left(\tens{I}:\tens{L}\right)-\tens{L}\vec{h}\right)\cdot \vec{n}\,da.
	\end{align}
	Application of the  divergence theorem gives:
	\begin{align}
		\mathcal{W}_{el}(\mathcal{V}(t)) = -\int_{\mathcal{V}(t)} \nabla\cdot \left(\Phi \left(\frac{D\vec{h}}{Dt}+\vec{h}\left(\tens{I}:\tens{L}\right)-\tens{L}\vec{h}\right)\right)\, dv. \label{app:Wel}
	\end{align}
\end{subequations}
By using Equations~(\ref{phicurrent})-(\ref{gausscurrent})
and defining $q = Q / J$ to be the total charge density in the current
configuration, we can establish the
following identities (adopting the summation convention):
\begin{subequations}
	\begin{align}
		\begin{aligned}
			\nabla \cdot \left(\frac{D\vec{h}}{Dt}\right)
			=
			\nabla\cdot \left(\frac{\partial \vec{h}}{\partial t}+ (\vec{v}\cdot \nabla)\vec{h}\right)
			=
			\frac{\partial \nabla\cdot \vec{h}}{\partial t}+ v_i\frac{\partial^2 h_j}{\partial x_i\partial x_j}\\
			+ \frac{\partial v_i}{\partial x_j} \frac{\partial h_j}{\partial x_i}
			=
			\frac{\partial q}{\partial t} + v_i  \frac{\partial}{\partial x_i}(\nabla \cdot \vec{h})+\nabla \vec{h}^T:\tens{L}
			=
			\frac{Dq}{Dt}+\nabla \vec{h}^T:\tens{L},
		\end{aligned}\\
		\begin{aligned}
			\vec{e} \tens{L} \vec{h}=  e_i L_{ij}h_j= e_ih_j L_{ij} = (\vec{e}\otimes \vec{h}):\tens{L},
		\end{aligned}\\
		\nabla\cdot(\tens{L}\vec{h})=
		\frac{\partial L_{ij}}{\partial x_i}h_j+ \frac{\partial h_{j}}{\partial x_i}L_{ij}
		=\frac{\partial }{\partial x_j}\left(\frac{\partial v_i}{\partial x_i}\right) h_j+\nabla \vec{h}^T:\tens{L}=\nabla(\tens{I}:\tens{L})\cdot \vec{h}
		+\nabla \vec{h}^T:\tens{L}.
	\end{align}
\end{subequations}
Thus, Equation \eqref{app:Wel} reduces to:
\begin{align}
	\mathcal{W}_{el}(\mathcal{V}(t))	= \int_{\mathcal{V}(t)} \left(\vec{e}\frac{D\vec{h}}{Dt}+\left[(\vec{e}\cdot\vec{h})\tens{I}-\vec{e}\otimes \vec{h}-\Phi(\nabla\cdot \vec{h})\tens{I}\right]:\tens{L}-\Phi \frac{Dq}{Dt}\right)\, dv,
\end{align} 

\section{Equilibrium solutions}\label{AppE}
\subsection{Constrained swelling}
Here we compute the homogeneous steady states of a gel in
equilibrium with an ionic bath for the case of two ionic species: $c_+$ and
$c_-$, with opposite charges, $z_+$ and $z_-$, respectively. 

In the following, we consider the ionic solution and the gel together, so we
introduce superscripts ${(b)}$ and ${(g)}$ to distinguish variables that correspond
to these two systems, respectively.
Starting from the ionic bath, we set all the temporal and spatial derivatives,
all fluxes and velocities in~(\ref{bath_model}) to zero. 

Let us assume that we can control the far-field concentration of the ions and 
impose $c_+^{(b)}\rightarrow c_0$. Then,
from the boundary conditions~(\ref{far_cond_bath}), we obtain that:
\begin{subequations} \label{final_Eq_bath}
	\begin{align}
		p^{(b)}\equiv 0,\quad
		c^{(b)}_+\equiv c_0,\quad
		\Phi^{(b)}\equiv 0,\quad
		c^{(b)}_-\equiv-\frac{z_+}{z_-} c_0, \quad	T^{(b)}\equiv 0.
	\end{align}
	Further assuming as in Yu et al.~\cite{Yu2017} that all species have the same
	characteristic molecular volume $\nu$, i.e. $\nu_s=\nu_+=\nu_-\equiv\nu$, the other equilibrium variables in the bath are defined by:
	\begin{align}
		\mu^{(b)}_s = \mu_s^0 + k_BT\ln \left(1-\left(1-\frac{z_+}{z_-}\right)\nu c_0 \right),\\[2mm]
		\mu^{(b)}_+ = \mu_+^0+k_BT\ln (\nu c_0),\\[2mm]
		\mu^{(b)}_- = \mu_-^0+k_BT\ln \left(-\frac{z_+}{z_-} \nu c_0\right).
	\end{align}
\end{subequations}
Analogously, we can also derive the set of algebraic equation for the 
homogeneous steady state in the gel from the system~(\ref{full_mod_gel}),
\begin{subequations}
	\begin{align}
		\tens{T}^{(b)}= -p^{(g)} + \frac{G\left(\lambda^2-1\right)}{J},\\
		0 = z_fc_f +z_+c^{(g)}_++z_-c^{(g)}_- \, \label{elecgel}\\
		\mu^{(g)}_s = p^{(g)} \nu + \mu_s^0+ k_BT\left[\ln (c^{(g)}_s \nu)+\frac{\chi(1-c^{(g)}_s\nu)+1}{J}\right], \\[2.5mm]
		\mu^{(g)}_\pm = p^{(g)} \nu + \mu^0_\pm +z_\pm e\Phi^{(g)} + k_BT \left[\ln (\nu c^{(g)}_\pm)-\frac{\chi c^{(g)}_s\nu-1}{J} \right],\\
		\lambda_0\lambda^2=J=\left(1-\nu\left(c^{(g)}_s+c^{(g)}_++c^{(g)}_-\right)\right)^{-1}.
	\end{align}\label{final_eq_gel}%
\end{subequations}
Using boundary conditions at the free interface, we can connect Equations~(\ref{final_Eq_bath}) and \eqref{final_eq_gel} by imposing continuity of chemical potentials and of the stress tensor in the direction normal to the interface. This leads to:
\begin{subequations}\label{equi1}
	\begin{align}
		-p^{(g)} \nu =  k_BT\left[\ln \left(\frac{\nu c^{(g)}_s}{1-\left(1-\frac{z_+}{z_-}\right)\nu c_0 }\right) +\frac{\chi(1-c^{(g)}_s\nu )+1}{J}\right],\label{museq} \\[2.5mm]
		-p^{(g)} \nu= z_+ e\Phi^{(g)}+ k_BT \left[\ln\frac{c^{(g)}_+}{c_0}-\frac{\chi c^{(g)}_s\nu-1}{J} \right],\label{mupeq}\\
		-p^{(g)} \nu= z_- e\Phi^{(g)}+ k_BT \left[\ln\left(-\frac{z_-c^{(g)}_-}{z_+c_0}\right)-\frac{\chi c^{(g)}_s\nu-1}{J} \right],\label{mumeq}\\
		-z_-c^{(g)}_- = z_fc_f +z_+c^{(g)}_+, \\
		p^{(g)}= \frac{G\left(\lambda^2-1\right)}{\lambda_0^2\lambda},\\
		\lambda_0^2\lambda=\left(1-\nu \left(c^{(g)}_s+c^{(g)}_++c^{(g)}_-\right)\right)^{-1}.
	\end{align}
\end{subequations}

Subtracting \eqref{mupeq} and \eqref{mumeq}, and subtracting ~\eqref{museq} from both~\eqref{mupeq} and~\eqref{mumeq}, we obtain the following relation between the electric potential $\Phi$ and the ionic concentrations:
\begin{subequations}
	\begin{align}
		c^{(g)}_-=-\frac{z_+}{z_-}\left[\kappa\right]^{1-\frac{z_-}{z_+}}\left(c^{(g)}_+\right)^{\frac{z_-}{z_+}},\label{DonA}\\
		\Phi^{(g)}= \frac{k_BT}{ez_+}\ln\left(\frac{\kappa}{c_s^{g}}\right),
	\end{align}
	where
	\begin{align}
		\kappa=\kappa\left(c^{(g)}_s,\lambda;\nu c_0,\lambda_0,\frac{z_+}{z_-}\right)=\exp\left(\frac{\chi}{\lambda_0^2\lambda}\right) \left[\frac{\nu c_0c_s^{(g)}}{1-\left(1-\frac{z_+}{z_-}\right)\nu c_0}\right].\label{kappaApp}
	\end{align}\label{mult}%
	Equation~(\ref{DonA}) is a generalisation of the standard Donnan Equilibrium \cite{donnan_theory_1924,huyghe_quadriphasic_1997} to our specific problem, where the additional exponential contribution in~(\ref{kappaApp}) due to the mixing energy has to be considered.
\end{subequations}
Using~\eqref{mult} to simplify the system~\eqref{equi1}, the latter reduces to:
\begin{subequations}
	\begin{align}\label{fin_eq}
		\mathcal{G}\frac{\left(\lambda^2-1\right)}{\lambda_0^2\lambda}+ \ln \left(\frac{\nu c^{(g)}_s}{1-\left(1-\zeta^{-1}\right)\nu c_0 }\right) +\frac{\chi(1-c^{(g)}_s\nu )+1}{\lambda_0^2\lambda}=0, \\[2.5mm]
		\left(c_+^{(g)}\right)^{1-\zeta}-\kappa^{1-\zeta}+\frac{\alpha_f}{\lambda_0^2\lambda}\left(c_+^{(g)}\right)^{-\zeta}=0,\label{fin_eq_2APP} \\
		\lambda_0^2\lambda-\left(1-\nu \left(c^{(g)}_s+c^{(g)}_+-\frac{\kappa^{1-\zeta}}{\zeta}\left(c^{(g)}_+\right)^{\zeta}\right)\right)^{-1}=0.
	\end{align}
	where $\mathcal{G}$ is a positive non-dimensional parameter corresponding to a scaled version of shear modulus, $\mathcal{G}=G\nu/(k_BT)$, $\alpha_f=z_f\nu C_f/z_+$ measures the number of fixed charges per molecule (relative to the valences of the fixed to mobile species), and $\zeta=z_-/z_+$. 
\end{subequations}
As discussed in the main text (see \S \ref{equil1D}), we can further simplify the above system by assuming $z_+=z_-$ to obtain Eq.~(\ref{constctext}).
\subsection{Free swelling}
\label{app:freeswlling}
For the free-swelling case the deformation tensor $\tens{F}=\lambda\tens{I}$ so that $\tens{B}=\lambda^2\tens{I}$. 
Similarly to \eqref{equi1} we obtain
\begin{subequations}
	\begin{gather}
		-p \nu =  k_BT\left[\ln \frac{\nu c^{(g)}_s}{1-2\nu c_0} +\frac{\chi(1-c^{(g)}_s\nu)}{J}\right],\label{old-temp4} \\[2.5mm]
		-p \nu= \pm e \Phi+ k_BT \left[\ln\frac{c^{(g)}_\pm}{c_0}-\frac{\chi c^{(g)}_s\nu}{J} \right],\label{old-temp5}\\
		c^{(g)}_- = z_fc_f +c^{(g)}_+, \\
		p = \frac{G\left(\lambda^2-1\right)}{\lambda^3},\label{old-pres}\\
		\lambda^3=\left(1-\nu\left(c^{(g)}_s+c^{(g)}_++c^{(g)}_-\right)\right)^{-1},
	\end{gather}
\end{subequations}
together with the relationship between the electric potential and the ionic concentrations:
\begin{equation}
\Phi= \frac{k_BT}{2e}\ln\left(\frac{c^{(g)}_-}{c^{(g)}_+}\right),
\end{equation}
while the concentrations $c_+$ and $c_-$ are given now by
\begin{equation}
c^{(g)}_\mp = \frac{c_0 c^{(g)}_s \nu}{1-2c_0\nu} \exp\left(\pm\frac{e\Phi}{k_BT}+\frac{\chi}{J}\right),
\end{equation}
which implies that
\begin{equation}
c^{(g)}_+c^{(g)}_-= \frac{c^2_0(c_s^{(g)})^2\nu^2}{\left(1-2c_0\nu\right)^2} \exp\left(\frac{2\chi}{J}\right).\label{old-mult}
\end{equation}
Combining the above with the electro-neutrality condition~(\ref{elecgel}), we can compute an expression for the equilibrium value of the ionic concentrations:
\begin{equation}
c^{(g)}_\pm = \mp \frac{z_fc_f}{2} + \sqrt{\left( \frac{z_fc_f}{2} \right)^2 + \frac{c^2_0(c^{(g)}_s)^2 \nu^2}{{\left(1-2c_0\nu\right)^2}} \exp\left(\frac{2\chi}{J}\right)},
\end{equation}
from which it can be seen that the equilibrium solution $\left(c^{(g)}_s,\lambda\right)=\left({c}^*_s,{\lambda^*}\right)$ is implicitly defined by the system of algebraic equations:
\begin{subequations}
	\begin{eqnarray}
	0=\mathcal{G^*}\frac{{\lambda^*}^2-1}{{\lambda^*}^3}+ \ln\left(\frac{{c}^*_s \nu}{1-2c_0\nu}\right) +
	\frac{\chi(1-{c}_s^*\nu)+1}{{\lambda^*}^3}\label{old-cond1}\\
	0=\nu{c}^*_s+2\sqrt{\left( \frac{\alpha_f}{2{\lambda^*}^3} \right)^2 + \frac{c^2_0{{c}^*_s}^2 \nu^2}{{\left(1-2c_0\nu\right)^2}} \exp\left(\frac{2\chi}{{\lambda^*}^3}\right)}-\frac{{\lambda^*}^3-1}{{\lambda^*}^3}\label{old-cond2},
	\end{eqnarray}
\end{subequations}
where $\mathcal{G}^*$ is the stiffness of the free-swelling gel.

\bibliographystyle{elsarticle-num-names} 
\bibliography{ref}
\end{document}